\documentclass[natbib]{svjour3}

\smartqed

\usepackage{graphicx}
\usepackage{color}           
\usepackage{url}    

 \usepackage{aps-bibstyle}  

\journalname{Space Science Reviews}

\newcommand{\aap}{{Astron. Astrophys.}}
\newcommand{\apj}{{Astrophys. J.}}
\newcommand{\jgr}{{J. Geophys. Res.}}

\newcommand{\solphys}{{Solar Phys.}}

\begin{document}

\title{Review article: MHD wave propagation near coronal null points of magnetic fields}

\titlerunning{MHD waves and coronal null points}

\author{J.~A. McLaughlin \and A.~W. Hood \and   I. De~Moortel  }

\authorrunning{McLaughlin {\it{et al.}}} 

\institute{J. A. McLaughlin \at
              School of Computing \& Engineering, Northumbria University, Newcastle Upon Tyne, NE1 8ST, UK \\
              \email{james.a.mclaughlin@northumbria.ac.uk}
}

\date{Received: 30 January 2010 / Accepted: 8 April 2010}  

\maketitle


\begin{abstract}
{
{We present a comprehensive review of MHD wave behaviour in the neighbourhood of coronal null points: locations where the magnetic field, and hence the local Alfv\'en speed, is zero.}
{The behaviour of all three MHD wave modes, i.e. the Alfv\'en wave and the fast and slow magnetoacoustic waves, has been investigated in the neighbourhood of 2D, 2.5D and {{(to a certain extent)}} 3D magnetic null points, for a variety of assumptions, configurations and geometries.}
{In general, it is found that the fast magnetoacoustic wave behaviour is dictated by the Alfv\'en-speed profile.  In a $\beta=0$ plasma, the fast wave is focused towards the null point by a refraction effect and all the wave energy, and thus current density,  accumulates close to the null point. Thus, {\emph{null points will be locations for preferential heating by fast waves}}.}
{Independently, the Alfv\'en wave is found to propagate along magnetic fieldlines and is confined to the fieldlines it is generated on. As the wave approaches the null point, it spreads out due to the diverging fieldlines. Eventually, the Alfv\'en wave accumulates along the separatrices (in 2D) or along the spine or fan-plane (in 3D). Hence, {\emph{Alfv\'en wave energy will be preferentially dissipated at these locations}}.}
{It is clear that the magnetic field plays a fundamental role in the propagation and properties of MHD waves in the neighbourhood of coronal null points. This topic is a fundamental plasma process and results so far have also lead to critical insights into reconnection, mode-coupling, quasi-periodic pulsations and phase-mixing.}
}
\end{abstract}


\section{Introduction}

Magnetohydrodynamic (MHD) wave motions (e.g.  Roberts \citeyear{Bernie};  Nakariakov \& Verwichte \citeyear{NV2005}; De Moortel \citeyear{DeMoortel2005}) are  ubiquitous throughout the solar corona (Tomczyk et al. \citeyear{Tomczyk}). Several different types of MHD wave motions have now been observed by various solar instruments:  slow magnetoacoustic  waves have been seen in {\emph{SOHO}} data (e.g. {{Ofman {\it{et al.}} \citeyear{Ofman1997}; DeForest \& Gurman \citeyear{plumes}}};  Berghmans \& Clette \citeyear{Berghmans1999}; Kliem {\it{et al.}} \citeyear{Kliem}; Wang {\it{et al.}} \citeyear{Wang2002}) and {\emph{TRACE}} data (De Moortel {\it{et al.}} \citeyear{DeMoortel2000}). Fast magnetoacoustic waves have been seen with {\emph{TRACE}} (Aschwanden {\it{et al.}} \citeyear{Aschwandenetal1999}, \citeyear{Aschwandenetal2002}; Nakariakov {\it{et al.}} \citeyear{Nakariakov1999}; Wang \& Solanki \citeyear{Wang2004}) and {\emph{Hinode}} (Ofman \& Wang \citeyear{OW2008}). Non-thermal line narrowing / broadening due to Alfv\'en waves has been reported by Harrison {\it{et al.}} (\citeyear{Harrison2002}) / {Erd{\'e}lyi} {\it{et al.}} (\citeyear{E1998}), {{Banerjee {\it{et al.}} (\citeyear{Banerjee1998}) }}and  O'Shea {\it{et al.}} (\citeyear{Oshea}). Alfv\'en waves have possibly been observed in the corona  (Okamoto {\it{et al.}} \citeyear{Okamoto}; Tomczyk {\it{et al.}} \citeyear{Tomczyk}) and chromosphere (De Pontieu {\it{et al.}} \citeyear{Bart2007}; Jess {\it{et al.}} \citeyear{Jess2009}), although these claims are subject to intense discussion (Erd{\'e}lyi \& Fedun \citeyear{RF2007}; Van Doorsselaere {\it{et al.}} \citeyear{Tom2008}).

It is clear that the  coronal magnetic field  plays a fundamental role in the  propagation and properties of MHD waves, and to begin to understand this  inhomogeneous, magnetised  environment, it is useful to look at the topology (structure) of the magnetic field itself.  Potential-field extrapolations of the coronal magnetic field can be made from photospheric magnetograms, and such extrapolations show the existence of important features of the topology: {\it{null points}} - locations where the magnetic field, and hence the Alfv\'en speed, is zero, and {\it{separatrices}} - topological features that separate regions of different magnetic flux connectivity. A comprehensive review can be found in Longcope (\citeyear{L2005}).

This paper will provide a comprehensive literature review of the nature of MHD wave propagation in the neighbourhood of coronal null points. This topic exists at the overlap of two important areas of solar physics: {{MHD wave and magnetic null-point theories}}. A brief introduction to both of these areas is provided in $\S\ref{section:MHDwaves}$ and $\S\ref{section:toplogy}$, including a mathematical description of magnetic null points. $\S\ref{section:polar}$ reviews the early work that considers a 2D null point in a cylindrically symmetric geometry and describes the system in terms of normal modes. $\S\ref{section:cartesian}$ reviews work performed in a 2D cartesian geometry that focuses on externally driven perturbations, and $\S\ref{section:nonlinear}$ describes the extension of these investigations into the nonlinear regime. $\S\ref{guide_field}$ details the effects of threading the 2D X-point with an orthogonal  weak-guiding field. $\S\ref{section:threedimensionalnullpoints}$ details the behaviour of MHD wave propagation in the neighbourhood of 3D null points, and the conclusions and summary are given in $\S\ref{section:conclusions}$.


\subsection{MHD Equations}\label{section:basic_MHD_equations}

The viscous, resistive, compressible  MHD equations {{utilised in this paper are}}:
\begin{eqnarray}
\rho \left[ {\partial {\bf{v}}\over \partial t} + \left( {\bf{v}}\cdot\nabla \right) {\bf{v}} \right] &=& - \nabla p + {\frac{1}{\mu}}\left(   { \nabla \times {\bf{B}}} \right)\times {\bf{B}} +  \nu  \nabla \cdot {\underline{\underline{\pi}}}  \;   \; ,\nonumber \\
 {\partial {\bf{B}}\over \partial t}  &=& \nabla \times \left ({\bf{v}}\times {\bf{B}}\right ) + \eta \nabla ^2  {\bf{B}}\;\; ,\nonumber \\
{\partial \rho\over \partial t} + \nabla \cdot \left (\rho {\bf{v}}\right ) &=& 0\; \;, \nonumber \\
 \rho \left[{\partial {\epsilon}\over \partial t}  + \left( {\bf{v}}\cdot\nabla \right) {\epsilon}\right] &=& - p \nabla \cdot {\bf{v}} + {{\frac{1}{\sigma}}} \left| {\bf{j}} \right| ^2 +  \nu \varepsilon_{ij} \pi_{ij}   \;\; \label{MHDequations}  ,
\end{eqnarray}
where $\rho$ is the mass density, ${\bf{v}}$ is the plasma velocity, ${\bf{B}}$ the magnetic induction (usually called the magnetic field), $p$ is the plasma pressure,  $ \mu = 4 \pi \times 10^{-7} \/\mathrm{Hm^{-1}}$  is the magnetic permeability,{{ $\nu$ is the coefficient of classical viscosity, $\pi_{ij} = \varepsilon_{ij} - \delta_{ij}\nabla \cdot {\bf{v}}$ is the stress tensor, $\varepsilon_{ij} = \left( {\partial v_i/ \partial x_j} + {\partial v_j/ \partial x_i} \right)/2$ is the rate-of-strain tensor,}} $\sigma$ is the electrical conductivity,  $\eta=1/ {\mu \sigma} $ is the magnetic diffusivity, $\epsilon= {p / \rho \left( \gamma -1 \right)}$ is the specific internal energy density, where $\gamma={5 / 3}$ is the ratio of specific heats and ${\bf{j}} = {{\nabla \times {\bf{B}}} / \mu}$ is the electric current density. $\nu$ and $\eta$ are assumed to be constants.

{{Note that the classical viscous term used in equations (\ref{MHDequations}) is in fact not the most appropriate for the solar corona since,  in the presence of strong magnetic fields, the viscosity takes the form of a non-isotropic tensor. However, only the papers of Craig \& Litvinenko (\citeyear{CL2007}) and Craig (\citeyear{Craig2008}) mentioned in this review  will invoke the non-isotropic viscous tensor and so, for brevity, we do not provide a full description here. The mathematical details of the  non-isotropic viscous tensor can be found in Braginskii (\citeyear{Braginskii1965}) and, for example, Van der Linden {\it{et al.}} (\citeyear{Linden}), Ofman {\it{et al.}} (\citeyear{Ofman1994}) and  Erd{\'e}lyi \& Goossens (\citeyear{EG1994}; \citeyear{EG1995}).}}


\subsection{MHD waves}\label{section:MHDwaves}

{{A wave is a disturbance that propagates through space and time, usually with the transference of energy. Such a disturbance, either continuous or transient, propagates by virtue of the elastic nature of the medium.}} In MHD, the magnetic tension provides an elastic restoring force, such that we would expect waves to propagate along uniform magnetic fieldlines with a characteristic speed:
\begin{eqnarray*}
v_A= \frac{ | {\bf{B}} |}{ \sqrt{ \mu \rho}} \;\;,
\end{eqnarray*}
where  $v_A$ is called the {\emph{Alfv\'en speed}}. Transverse waves travelling at this speed along magnetic fieldlines are called {\emph{Alfv\'en waves}}.

If we consider a compressible medium, then we can define the {\emph{sound speed}} as: 
\begin{eqnarray*}
c_s  = \sqrt{\frac{\gamma p}{\rho}} \;\;. 
\end{eqnarray*}
When assuming  a compressible medium, the Alfv\'en wave still remains, but the sound and Alfv\'en speed can now couple together to give {\emph{magnetoacoustic}} waves. Two combinations arise: the higher frequency mode is known as the {\emph{fast magnetoacoustic wave}} and the lower frequency wave is known as the {\emph{slow magnetoacoustic wave}}. These three wave types, the Alfv\'en wave and the fast and slow magnetoacoustic waves, make up the three MHD waves considered in this review paper.

The fundamental properties and nature of linear MHD waves in uniform magnetic fields have been reported in detail by several authors, for example in an unbounded homogeneous medium (Cowling \citeyear{Cowling1976}), and in a bounded inhomogeneous slab / cylindrical density profile embedded in a uniform magnetic field (Roberts \citeyear{Roberts1981a}) / (Edwin \& Roberts \citeyear{ER1983}; Cally \citeyear{Cally1986}; Roberts \& Nakariakov \citeyear{RN2003}).

Finally, in MHD it is important to consider the ratio of magnetic pressure to thermal pressure. This ratio is called the plasma$-\beta$ and is given by:
\begin{eqnarray}
\beta=  \frac{2 \mu p}{| {\bf{B}} |^2} = {\frac{2}{\gamma}}{\frac{c_s^2}{v_A^2}}\;\;.\label{plasmabetaequation}
\end{eqnarray}
The properties of the fast and slow  magnetoacoustic waves have a strong dependence on the magnitude of the plasma$-\beta$, namely because it is directly proportional to the square of the ratio of the sound speed to the Alfv\'en speed. Thus, in a regime where $\beta \ll 1$, magnetic pressure and magnetic tension dominate the propagation and vice versa. Table 1 lists the main properties of the three wave types depending upon their environment. Note that the Alfv\'en wave behaviour is independent of the plasma-$\beta$, as it is a purely magnetic wave (in the linear regime).

The plasma$-\beta$ parameter varies greatly with height across the layers of the solar atmosphere (see Gary \citeyear{Gary2001} for well-constrained values). However, magnetic pressure generally dominates thermal pressure in the solar corona, and thus it is usual to assume plasma$-\beta \ll 1$ when modelling a coronal environment. Hence, when we talk about coronal null points, we are talking about null points in a low$-$ or zero$-\beta$ environment, although there are some caveats to this ($\S\ref{section:mode conversion}$).

\begin{table}[h]
\begin{tabular}{|c|c|c|}
\hline
     & plasma-$\beta\gg1$ (high-$\beta$) & plasma$-\beta\ll 1$ (low-$\beta$)\\
\hline\hline
 Alfv\'en wave & \multicolumn{2}{c}{Transverse wave propagating at speed $v_A$}
\begin{tabular}{c}
\end{tabular} \vline\\
\hline
Fast MA wave & \begin{tabular}{c}
\vspace{0.1cm}Behaves like sound wave\\
 (speed $c_s$)
\end{tabular} & \begin{tabular}{c}
Propagates roughly isotropically \\ Propagates across magnetic fieldlines \\
(speed $v_A$) \\
\end{tabular}\\
\hline
Slow MA wave & \begin{tabular}{c}
Guided along ${\bf{B}}$  \\
(speed $v_A$)
\end{tabular} & \begin{tabular}{c}
Guided along ${\bf{B}}$\\
Longitudinal wave propagating \\
at speed $c_s$\\
\end{tabular}\\
\hline
\end{tabular}
\caption{Properties of MHD waves depending on the plasma$-\beta$.}
\end{table}


\subsection{Magnetic Topology}\label{section:toplogy}

The magnetic field plays an essential role in understanding the myriad of phenomena in the solar corona. A realistic magnetic field can have many different components, and we can use {\emph{topology}} nomenclature to reduce a complicated set of fieldlines to something more understandable. In 2D, a general magnetic configuration contains {\emph{separatrix curves}} (separatrices) which split the magnetic plane into topologically distinct regions, in the sense that within a specific region all the fieldlines start at a particular source and end at a particular sink. There is a second important topological aspect: {\emph{magnetic null points}} (or neutral points) are single-point locations where the magnetic field vanishes (${\bf{B}}={\bf{0}}$). There are two types of magnetic null point: {\emph{X-type null points}}, commonly called X-points, which  occur at the intersection of separatrix curves, and {\emph{O-type null points}}, or O-points, located at the center of magnetic islands. Magnetic topologies that contain null points are common in the presence of multiple magnetic sources.

A magnetic fieldline that joins two null points (itself a special type of separatrix) is called a {\emph{separator}}. Thus, instead of showing all the magnetic field lines in a region, we can just show the important aspects of the topology; such a picture of the magnetic structure is called the {\emph{magnetic skeleton}} of the field. In 3D, we have similar properties, now with {\emph{separatrix surfaces}} separating the volume into topologically distinct regions, and these surfaces intercept at a separator.


\subsubsection{Mathematical description of null points}\label{2Dnulls}


\begin{figure}[t]
\includegraphics[width=6.0in]{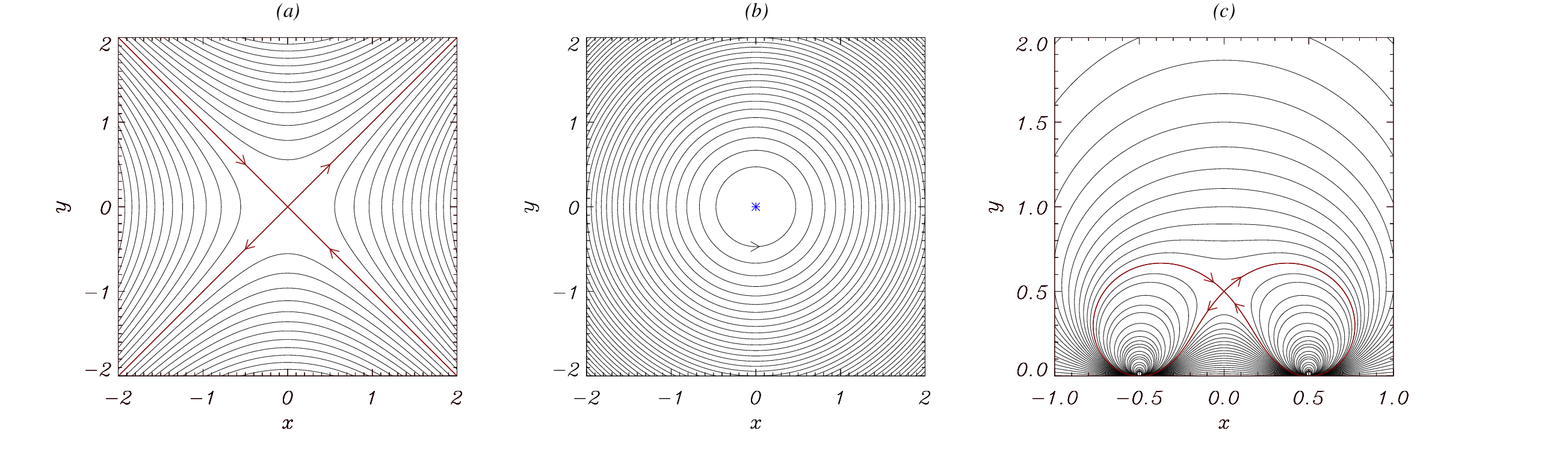}
\caption{$(a)$ X-type null point for $\alpha^2=1$. This potential neutral point has separatrices (red lines) intersecting at an angle of $\pi/2$. $(b)$ O-type null point, with $\alpha =-1$. A blue star denotes the null point. $(c)$ Single potential magnetic null point configuration created by interaction of two dipoles. Here, $A_z= y/ \left[ \left( x+\lambda \right)^2 +y^2\right] + y/ \left[ \left( x-\lambda\right)^2 + y^2 \right]$, for $\lambda=0.5$. Red lines denote the separatrices.}
\label{figure1}
\end{figure}


Let us first consider null points in 2D {{(e.g. Dungey \citeyear{Dungey1953}; \citeyear{Dungey1958}). Following  $\S1.3.1$ of Priest \& Forbes ({\citeyear{magneticreconnection2000}})}}, we assume a magnetic field of the form:
\begin{eqnarray*}
{\bf{B}}= \left[ B_x(x,y), B_y(x,y),0\right]\;\;.
\end{eqnarray*}
A null point occurs at the point $(x_0,y_0)$ if:
\begin{eqnarray*}
B_x(x_0,y_0) =0\;\;{\rm{and}}\;\;B_y(x_0,y_0) =0\;.
\end{eqnarray*}
Expanding $B_x$ and $B_y$ in a Taylor series about $(x_0,y_0)$ gives the linear approximation:
\begin{eqnarray}
B_x &=& \left.{\frac{\partial B_x}{\partial x}}\right|_{(x_0,y_0)} \left(x-x_0\right) + \left.{\frac{\partial B_x}{\partial y}}\right|_{(x_0,y_0)} \left(y-y_0\right)\nonumber \\
 &=& a(x-x_0)+b(y-y_0)\;\;,\label{a1}\\
B_y &=& \left.{\frac{\partial B_y}{\partial x}}\right|_{(x_0,y_0)} \left(x-x_0\right) + \left.{\frac{\partial B_y}{\partial y}}\right|_{(x_0,y_0)} \left(y-y_0\right) \nonumber \\
&=& c(x-x_0)-a(y-y_0)\;\;,\label{a2}
\end{eqnarray}
where the coefficients $a,b,c$ are arbitrary.

Let us now introduce the vector potential (also called the flux function), ${\bf{A}}$, such that ${\bf{B}} = \nabla \times {\bf{A}}$, and in 2D we have ${\bf{A}} = (0,0,A_z)$. Thus, we have: 
\begin{eqnarray}
{\bf{B}}= \left(\frac{\partial A_z}{\partial y}, -\frac{\partial A_z}{\partial x},0\right)\;\;.\label{fluxfunction}
\end{eqnarray}

Integrating equations (\ref{a1}) and (\ref{a2}) gives the corresponding vector potential as:
\begin{eqnarray*}
A_z= a(x-x_0)(y-y_0)+ \frac{b}{2} (y-y_0)^2 - \frac{c}{2}(x-x_0)^2\;\;,
\end{eqnarray*}
where we have chosen the arbitrary constant of integration such that $A_z$ vanishes at  $(x_0,y_0)$. Further simplification is possible by rotating the $xy$-axes through an angle $\theta$ to give new $x'$, $y'$-axes, and choosing the angle $\theta$ such that $\tan{2\theta}= -2a/(b+c)$. This simplification gives the corresponding vector potential as:
\begin{eqnarray}
A_z= {\frac{B}{2L}}\left[ \left(y'-y'_0\right)^2 - \alpha^2 \left(x'-x'_0\right)^2\right]\;\;,\label{ff_simpleXpoint}
\end{eqnarray}
where
\begin{eqnarray*}
{\frac{B}{L}}= \frac{2a^2+b^2-c^2}{\sqrt{4a^2+(b+c)^2}}\;\;,\;\; \alpha ^2 = {\frac{ 4a^2}{ 2a^2+b^2-c^2} -1} \;\;.
\end{eqnarray*}

Here, $B$ is the characteristic strength of the magnetic field and $L$ is the characteristic length-scale over which the field varies. The corresponding field components are:
\begin{eqnarray}
B_x =\frac{B}{L}(y'-y'_0)  \;\; {\rm{and}} \;\; B_y =\frac{B}{L}\alpha^2 (x'-x'_0) \label{BX_and_BY}\;\;.
\end{eqnarray}

Magnetic field lines are defined by $A_z$ equal to a constant. For $\alpha^2>0$, the fieldlines are hyperbolic, giving an X-type null point. The separatrices are given by $y'-y'_0 = \pm \alpha (x'-x'_0)$ and are inclined at an angle $\pm \arctan{\alpha}$ to the $x'$-axis.  The magnetic fieldlines for $\alpha^2=1$, $x'_0=y'_0=0$ can be seen in  Figure \ref{figure1}a.


\begin{figure*}
\includegraphics[width=6.0in]{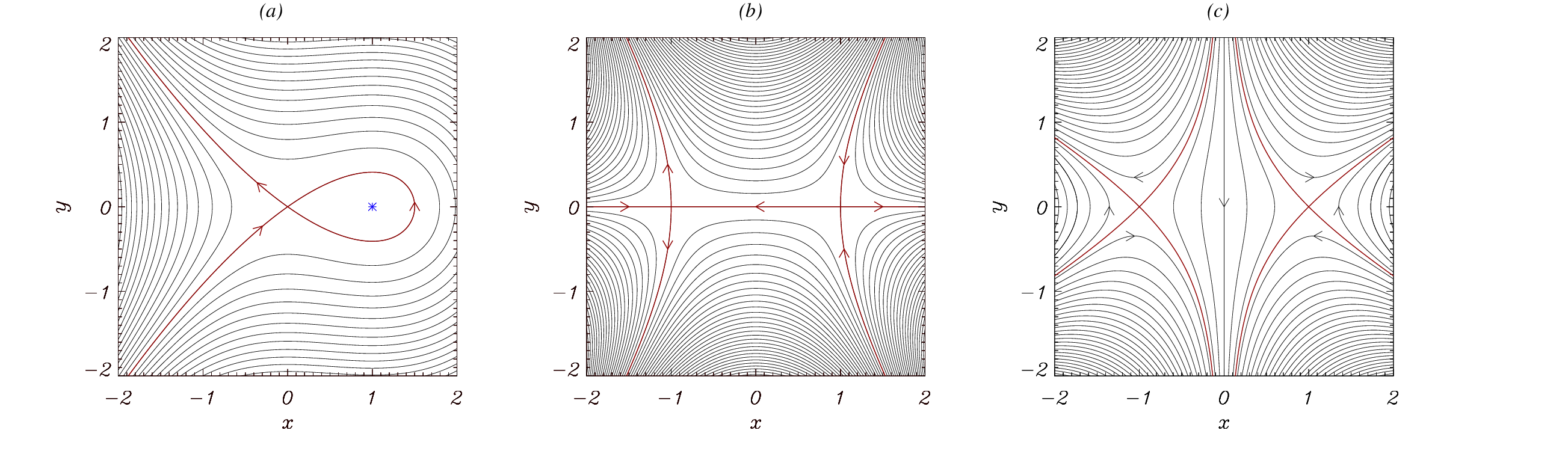} 
\caption{Magnetic fields containing two null points. $(a)$ Magnetic configuration containing both X-type and O-type null points. Here $A_z= \lambda^2 x- y^2 - (x-\lambda)^3/3$ (${\bf{B}}=\left[-2y, \left(x-\lambda\right)^2 - \lambda^2\right]$) for $\lambda=0.5$. Red lines/blue star denotes the separatrices/O-type null point. $(b)$ Potential magnetic configuration containing two X-type null points connected by a separator. Here, $A_z= -x^2y+y^3/3+ \lambda^2 y$, where $\lambda=1$.  $(b)$ Potential magnetic configuration containing two X-type null points not connected by a separator. Here, $A_z= -xy^2+x^3/3-\lambda^2 x$, where $\lambda=1$. Red lines denote the separatrices.}
\label{figure2}
\end{figure*}


The value of $\alpha$ (and thus the angle between the separatrices) is related to the current density. The current density is given by:
\begin{eqnarray*}
{\bf{j}}= {\frac{1}{\mu}}\left( {\nabla \times {\bf{B}}}\right) = -{\frac{1}{\mu}} \nabla^2 A_z  {\hat{\bf{z}}} =  -{\frac{B}{\mu L}} \left( 1-\alpha^2 \right) {\hat{\bf{z}}}  \;\;.
\end{eqnarray*}
Thus, a null point is potential if $\alpha=\pm1$ (i.e. an X-type configuration of rectangular hyperbola) and the angle between the separatrices is $\pi/2$. Note that for an O-point, $\alpha$ is imaginary and so the current density is always non-zero. Thus, O-type neutral points can never be potential.

An O-type null point magnetic configuration can be seen in Figure \ref{figure1}b, for $\alpha^2=-1$, $x'_0=y'_0=0$. However, note that the simple 2D magnetic field configuration of equation (\ref{BX_and_BY}) is only valid close to the null point: as $x'$ and/or $y'$ get very large, ${\bf{B}}$ becomes unphysically large. Figure \ref{figure1}c denotes a more realistic single magnetic null point configuration created by the interaction of two dipoles. This configuration comprises of four separatrices and an X-point, and as $x'$ and/or $y'$ get large, the field strength becomes smaller (i.e. a more physical field).

Magnetic configurations can also contain multiple null points, and it can be argued that null points appear in pairs; a double null point may arise as a local  bifurcation of a single 2D null point (see e.g. Galsgaard {\it{et al.}} \citeyear{KRRN1996}; Brown \& Priest \citeyear{BP1998}).    Figure \ref{figure2}a.  shows a magnetic configuration containing both X-type and O-type null points. Figures \ref{figure2}b and \ref{figure2}c present magnetic configuration containing two X-type null points connected by a separator and not connected by a separator, respectively.


\begin{figure}
\hspace{-1.0cm}
\includegraphics[width=6.0in]{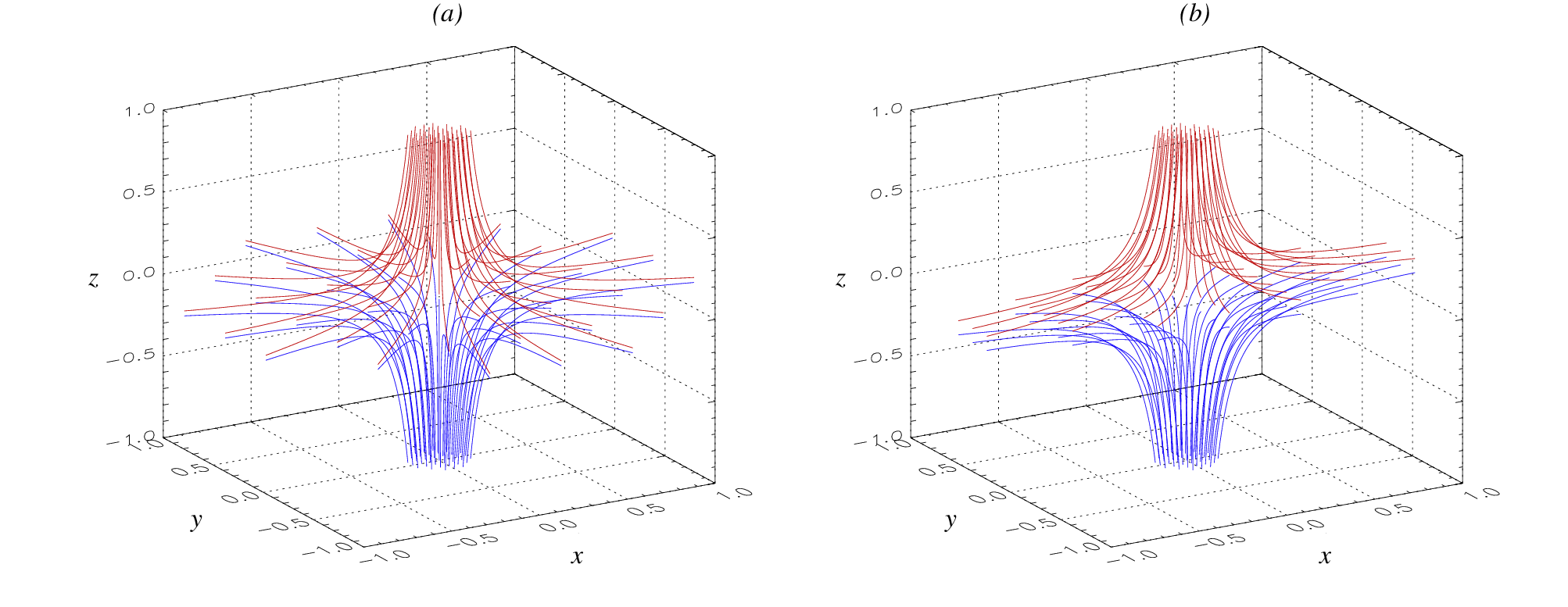}
\caption{$(a)$ Proper radial null point, described by ${\bf{B}}=(x,y,-2z)$, {\it{i.e.}} $\epsilon=1$. $(b)$ Improper radial null point, described by ${\bf{B}}=\left(x,\epsilon y,- \left[\epsilon+1 \right]z\right)$, for $\epsilon={1}/{2}$.  Note for $\epsilon={1}/{2}$, the field lines rapidly curve such that they run parallel to the $x-$axis along $y=0$. In both figures, the  $z-$axis indicates the {\emph{spine}} and the $xy-$plane at $z=0$ denotes the {\emph{fan-plane}}. The red fieldlines have been tracked from the $z=1$ plane, the blue from $z=-1$.}
\label{figure3}
\end{figure}


\subsubsection{Three-dimensional magnetic null points}\label{section:threeDnulls}

Magnetic null points also exist in three dimensions, but  occur in a different form to those described in \S\ref{2Dnulls}. In 3D, potential null points are of the form:
\begin{eqnarray}
{\bf{B}} = \frac{B}{L} \left(x,\epsilon y,-\left[ \epsilon +1 \right] z \right)   \label{Bfield}  \;\;,
\end{eqnarray}
where the parameter $\epsilon$ is related to the predominate direction of alignment of the fieldlines in the fan plane.  Parnell {\it{et al.}} (\citeyear{Parnell1996}) investigated and classified the different types of linear magnetic null points that can exist (our $\epsilon$ parameter is called $p$ in their work). Topologically, this 3D null consists of two key parts: the $z-$axis represents a special, isolated fieldline called the {\emph{spine}} which approaches the null from above and below (as found by Priest \& Titov \citeyear{PriestTitov1996}) and the $xy-$plane through $z=0$ is known as the {\emph{fan-plane}} and consists of a surface of fieldlines spreading out radially from the null. Figure \ref{figure3} shows two examples of 3D null points: $\epsilon=1$ (Figure \ref{figure3}a) and  $\epsilon=1/2$ (Figure \ref{figure3}b). Titov \& Hornig (\citeyear{TH2000}) have investigated the steady state structures of 3D magnetic null points.

Equation (\ref{Bfield}) is the general expression for the linear field about a potential magnetic null point (see Parnell {\it{et al.}} \citeyear{Parnell1996}: ${\S}{IV}$).  For $\epsilon \ge 0$, 3D nulls are described as {\it{positive}} nulls, {\it{i.e.}} the spine points into the null and the field lines in the fan are directed away. In addition, all potential nulls are designated  {\it{radial}}, {\it{i.e.}} there is no spiral motions in the  fan-plane.

In this review paper, we only consider positive, potential null points, and thus there are three cases to consider:
\begin{itemize}
\item{$\epsilon  = 1$: describes a {\it{proper null}} (Figure \ref{figure3}a). This magnetic null has cylindrical symmetry about the spine axis (so is actually only a 2.5D null point).}
\item{$\epsilon > 0,\: \epsilon \neq 1$: describes an {\it{improper null}}  (Figure \ref{figure3}b). Field lines rapidly curve such that they run parallel to the $x-$axis if $0<\epsilon <1$ and parallel to the $y-$axis if $\epsilon >1$.}
\item{$\epsilon=0$: equation (\ref{Bfield}) reduces to a simple 2D X-point potential field in the $xz-$plane and forms a null line along the $y-$axis through $x=z=0$.}
\end{itemize}


\subsection{Statistics of coronal null points}\label{statistics}

We have provided a mathematical description of null points, but how common are null points in the corona? Null points are an inevitable consequence of the distributed isolated magnetic flux sources at the photospheric surface. {{Using photospheric magnetograms to provide the field distribution on the lower boundary,  both potential and non-potential (nonlinear force-free) field extrapolations suggest}} that there are always likely to be null points in the corona. The number of such singular points will depend upon the magnetic complexity of the photospheric flux distribution. Detailed investigations of the coronal magnetic field, using such potential field calculations, can be found in {Beveridge} {\it{et al.}} (\citeyear{Beveridge2002}) and {Brown \& Priest (\citeyear{BrownPriest2001})}. The properties of coronal null points have also been considered through theoretical considerations (e.g. Parnell {\it{et al.}} \citeyear{Parnell1996}; Brown \& Priest \citeyear{BrownPriest2001};  {Beveridge} {\it{et al.}} \citeyear{Beveridge2002}; Parnell \& Galsgaard \citeyear{PG2004}; Parnell {\it{et al.}} \citeyear{Parnell2008}).

The statistics of coronal null points has been investigated  using two methodologies: direct measurement from potential field extrapolations (e.g. Close {\it{et al.}} \citeyear{Close2004}; R\'egnier {\it{et al.}} \citeyear{Stephane2008}) and, secondly, as an estimate from the Fourier spectrum of magnetograms (Longcope \& Parnell \citeyear{LP2009}). Close {\it{et al.}} (\citeyear{Close2004}) calculated a potential field extrapolation from a high resolution MDI magnetogram and found $1.7\times10^{-3}$  magnetic null points per square megameter. R\'egnier {\it{et al.}} (\citeyear{Stephane2008}) performed a similar investigation using a magnetogram from the Narrowband Filter Imager onboard {\emph{Hinode}} and found  $6.7\times10^{-3}$ ${\rm{Mm^{-2}}}$. Longcope \& Parnell (\citeyear{LP2009}) investigated 562 MDI magnetograms using the Fourier spectrum of magnetograms and found $3.1\times10^{-3}\pm 3.0\times10^{-4}$  coronal  null points per square megameter (at altitudes greater than 1.5 Mm). Alternatively, we can estimate the total number of coronal null points by multiplying these results by the surface area of the Sun (i.e. to provide a crude estimate, where we assume the Sun is free of active regions and coronal holes). This corresponds to approximately $10,000$  (Close {\it{et al.}} \citeyear{Close2004}),  $19,000$  (Longcope \& Parnell \citeyear{LP2009}) or $40,000$  (R\'egnier {\it{et al.}} \citeyear{Stephane2008}) coronal null points.

{{More recently,  Cook {\it{et al.}} (\citeyear{Cook2009}) investigated the solar cycle variation of coronal null points using a potential field source surface model in spherical geometry, and find that there is no significant variation in the number of nulls found from the rising to the declining phase (indicating that null points are present throughout both phases of the solar cycle).}}

{{Several investigations also consider specific examples of null points in the corona. For example, Aulanier {\it{et al.}} (\citeyear{Aulanier}) investigated a class M3 flare that occured on 14 July 1998 above a $\delta-$spot. Using potential field extrapolations, the authors recreated the pre-flare magnetic topology  from  Kitt Peak line-of-sight magnetograms and revealled a single coronal null point located above the $\delta-$spot. Secondly, Ugarte-Urra {\it{et al.}} (\citeyear{Ugarte}) investigated the magnetic topology of  26 CME events by performing potential field extrapolations from corresponding MDI magnetograms, and find that magnetic null points are present in a large number of the pre-CME topologies.}}

Finally, we note that a null point plays a key role in the {\emph{magnetic breakout model}} (e.g. Antiochos \citeyear{Antiochos1998}; Antiochos {\it{et al.}} \citeyear{Antiochos1999}; MacNeice {\it{et al.}} \citeyear{Macneice2004}; Lynch {\it{et al.}} \citeyear{lynch2004}; Choe {\it{et al.}} \citeyear{choe2005}). The equilibrium set-up of the  magnetic breakout model  model consists of a quadrupolar photospheric flux distribution coupled with an overlying field, and such a set-up contains a coronal null point. Such a null point is an ideal candidate for the study of MHD wave behaviour about coronal null points (i.e. the investigations detailed in this review paper). However, it is important to stress  that the null point in the breakout model is not the only candidate - the ideas and investigations detailed below apply equally well around coronal null points found elsewhere (i.e. quiet Sun and inside active regions). Thus,  the coronal null points we are describing in this paper are not solely those involved in the magnetic breakout model.


\subsection{Why is this area of study interesting or important?}\label{section:motivation}

{{The motivation for investigating the behaviour of MHD wave propagation in the neighbourhood of magnetic null points can be summarised as follows:
\begin{itemize}
\item{MHD wave propagation in inhomogeneous media is a fundamental plasma process, and the study of MHD wave behaviour in the neighbourhood of magnetic null points directly contributes to this area.}
\item{We now know that MHD wave perturbations are omnipresent in the corona. We also know that null points are an inevitable consequence of the distributed isolated magnetic flux sources at the photospheric surface (where the number of such singular points will depend upon the magnetic complexity of the photospheric flux distribution). Thus, these two areas of scientific study; MHD waves and magnetic topology, {\emph{will}} encounter each other at some point,  i.e. MHD waves will propagate into the neighbourhood of coronal null points (e.g. blast waves from a flare will at some point encounter a null point). Thus, the study of MHD waves around null points is itself a fundamental coronal process.}
\item{The study of MHD wave behaviour in the neighbourhood of magnetic null points is also interesting in its own right and, as we shall see, often  provides critical insights into other areas of plasma physics, including: mode-conversion ($\S\ref{section:mode conversion}$), reconnection ($\S\ref{section:nonlinear}$), quasi-periodic pulsations ($\S\ref{QPPs_section}$) and  phase-mixing ($\S\ref{PM_section}$).}
\end{itemize}

}}


\section{Two-dimensional null points in a cylindrically symmetric geometry}\label{section:polar}


The first  investigation into the behaviour of MHD waves in the neighbourhood of 2D magnetic null point  was performed by  Bulanov \& Syrovatskii (\citeyear{Bulanov1980}). The authors  considered the linearised MHD equations (equations \ref{MHDequations}) under the cold plasma approximation ($\beta=0$) in a cylindrically symmetric geometry, where a circular boundary is imposed at $r=1$ at which the fieldlines are held fixed. Their  investigation produces a detailed discussion of the propagation of fast and Alfv\'en waves in the neigbourhood of an equilibrium magnetic field:
\begin{eqnarray}
 {\bf{B}}_0 = \frac{B}{L} \left( -y, -x \right)\;\;,
\end{eqnarray}
which corresponds to the magnetic configuration in Figure \ref{figure1}a. The subscript $0$ denotes an equilibrium quantity.

 Bulanov \&  Syrovatskii also noted that in this 2D geometry, the  $z-$component motions are decoupled from $xy-$plane motions.  Thus, the Alfv\'en wave, which is governed by motions transverse to the magnetic field (i.e. $v_z \: {\hat{\bf{z}}}$) is decoupled from the magnetoacoustic waves (motions in the  $xy-$plane). Hence, it is possible to consider the  Alfv\'en wave and the magnetoacoustic waves separately in a 2D geometry.


In their paper, harmonic fast waves are generated at the $r=1$ boundary and these propagate inwards towards the null point, and Bulanov \& Syrovatskii find that, in the asymptotic limit $r\to 0$, the  perturbations have  azimuthal symmetry, i.e. propagate as cylindrical waves. This was the first piece of work that indicated a key relationship between fast waves and null points. However, the assumed cylindrical symmetry means that the disturbances can only propagate either towards or away from the null point, and are already encircling the null point. Thus, it is unclear if this is a general result.

To investigate the Alfv\'en wave,  Bulanov \& Syrovatskii  make a coordinate transformation such that the coordinate lines coincide with the lines of force, where:
\begin{eqnarray}
 \zeta = \frac{1}{2}\left(x^2-y^2\right) = \frac{1}{2}{r^2\cos{2\theta}} \;\;, \eta = xy = \frac{1}{2}{r^2\sin{2\theta}}\;\;.\label{BS_transform}
\end{eqnarray}
The authors find that Alfv\'en perturbations propagate along magnetic fieldlines at the local Alfv\'en speed.   The inhomogeneity of the Alfv\'en speed profile leads to an exponential increase in the gradients in the system, and in the asymptotic limit $t \to \infty$, these gradients accumulate at the separatrices of the field, i.e. $\zeta=0$. Again, this result was the first to indicate a relationship between Alfv\'en wave propagation and the location of the separatrices.

An alternative approach to the study of MHD wave behaviour in the neighbourhood of a 2D magnetic null point  was investigated in a series of papers by  Craig and co-workers (Craig \& McClymont \citeyear{CM1991};  \citeyear{CM1993}; Craig \& Watson \citeyear{CW1992}) in which the authors considered perturbations of the flux function $A_z$. Here, the focus was on applications for magnetic reconnection, specifically to  investigate if null points (viewed as weaknesses in the magnetic field) could collapse in response to imposed boundary motions. To clearly demonstrate their results, we repeat part of their analysis here:



In polar coordinates, an X-type null point, located at the origin, can be expressed as:
\begin{eqnarray}
{\bf{B}}_0 = \frac{B}{R} \left(r \sin{2 \theta} \:{\hat{\bf{r}}} + r \cos{2 \theta}\: {\bf{\hat{\bf{\theta}}}}  \right)  \;\;,\label{simple_X_point_polar}
\end{eqnarray}
where ${\bf{B}}_0$ denotes the equilibrium magnetic field and  $R$ represents the typical size of a coronal magnetic structure.

The MHD equations (\ref{MHDequations}) can be simplified by invoking the flux function $A_z$ (equation \ref{fluxfunction}). In polar coordinates, the equilibrium flux function ($A_0$) for a simple 2D X-point (i.e equation \ref{ff_simpleXpoint}) is:
\begin{eqnarray}
A_0=-\frac{1}{2} r^2 \cos{2\theta} \label{A0_equation}
\end{eqnarray}
where the evolution of $A_z$ is governed by the induction equation and the dynamics are governed by the momentum equation:
\begin{eqnarray*}
\frac{\partial {{A}_z}}{\partial t} +  \left( {\bf{v}}\cdot\nabla \right) {{A_z}} &=& \eta \nabla ^2  {{A_z}}\; \;,\\
\frac{\partial {\bf{v}}} {\partial t} + \left( {\bf{v}}\cdot\nabla \right) {\bf{v}} &=& - \nabla A_z \cdot \nabla^2 A_z\;\;.
\end{eqnarray*}
These equations can be linearised about the equilibrium flux function such that $A_z= A_0 + A_1$, and combined to form a single differential equation for the perturbed flux function ($A_1$):
\begin{eqnarray}
\frac{\partial^2 A_1}{\partial t^2}  = | \nabla A_0 |^2 \nabla^2 {A_1}  + \eta \nabla^2 {\frac{\partial A_1}{\partial t}}  \label{craig}\;\;,
\end{eqnarray}
where, from equation (\ref{A0_equation}), $| \nabla A_0 |^2=r^2$.

We can see that the right-hand-side of equation (\ref{craig}) has two parts. Consider a region close to the origin, $r_c$, where $r_c\sim\eta^{1/2}$ is the skin depth. Far from the origin, $r\gg r_c$, advection effects dominate and equation (\ref{craig}) reduces to a wave equation:
\begin{eqnarray}
\frac{\partial^2 A_1}{\partial t^2}  = r^2 \nabla^2 {A_1}\label{focusss}
\end{eqnarray}
Here, the rate of propagation of information is governed by the wave speed proportional to $r$ which makes the signal travel time logarithmic in $r$. Thus, a disturbance on the outer boundary ($r=1$) propagates into the diffusion region in a time $\delta t \sim | \ln {r_c} |\sim\frac{1}{2} \ln{ \eta }$.

Alternatively, close to the origin, $r\ll r_c$, and equation (\ref{craig}) reduces to the diffusion equation:
\begin{eqnarray*}
\frac{\partial A_1}{\partial t}  = \eta \nabla^2 A_1\;\;.
\end{eqnarray*}
Diffusion is ultimately responsible for dissipating the kinetic and magnetic energy in the system.

Next, let us assume separation of variables such that:
\begin{eqnarray*}
A_1(r,\theta,t) = f(r)  e^{i m \theta}  e^{\lambda t}   \;\;,
\end{eqnarray*}
where $m$ is the azimuthal wavenumber. The eigenequation for $f(r)$ is then:
\begin{eqnarray}
r \left[\frac{\partial}{\partial r} \left( r \frac{\partial f}{\partial r}\right) \right] = \left[ \frac{\lambda^2}{\left(1+\eta \lambda / r^2\right)} + m^2\right] f  \;\;.\label{f_r}
\end{eqnarray}

Close to the origin ($r\ll r_c$), the radial dependence of the current $j_z = -\nabla^2 A_1=  J_m \left[ \left( - \lambda / \eta \right)^{1/2} r \right]$. We note that only the $J_0$ Bessel function is nonzero at the origin (which is equivalent to a nonvanishing displacement in the flux function) and so the $m=0$ mode is the   only mode which allows topological reconnection. Assuming $m=0$, the evolution of $A_z$ can then be obtained  from the analytical or numerical solution of equation (\ref{craig}).


Craig \& McClymont (\citeyear{CM1991}) derived  equation (\ref{craig}) and investigated the fieldlines passing through the (circular) boundary in a particular manner in order to perturb the field - shifting the footpoints so as to \lq\lq{close up}\rq\rq{} the angle of the X-point.  The resulting field perturbations cause the null point to collapse to form a current sheet in which reconnection can release magnetic energy. In these models the boundary motions move the field lines but do not return them to their original positions (akin to photospheric footpoint motion). Thus, the Poynting flux induced by the imposed motion (and then fixing the field after the motion is complete) accumulates at the resulting current sheet and provides the energy released in the reconnection.

Craig \& McClymont find that, as their system evolves, the fieldlines reconnect as they pass through the null point (located at the origin), indicating that resistivity is essential to this mode. Their initial vertical \lq{current sheet}\rq{ } begins reconnecting and the inertia of the flowing plasma carries the system past the equilibrium configuration, until a weaker horizontal current sheet is formed. Then, a much weaker vertical current sheet returns at a later time (one complete cycle). After three such cycles, the system is close to its equilibrium (potential) configuration. The reconnection is found to be {\emph{oscillatory}}, with inertial overshoot of the plasma carrying more magnetic flux through the neutral point than is required to reach a static equilibrium. The reconnection rate scales as $ | \ln {\eta} | ^2$, and so the reconnection is described as {\emph{fast}}. This is an important results as it shows that the damping of the fast wave is still highly significant even when the resistivity is small.

Craig \& Watson (\citeyear{CW1992}) considered the radial propagation of the $m=0$ mode and solved equation (\ref{f_r}) using a mixture of analytical and numerical solutions. They demonstrated that the propagation of the $m=0$ wave towards the null point generates an exponentially large increase in the current density and that magnetic resistivity dissipates this current in a time related to $\log { \eta }$, in agreement with Craig \& McClymont (\citeyear{CM1991}). Their initial disturbance is given as a function of radius, i.e. an internal perturbation is considered. In their investigation, the outer radial boundary is held fixed so that any outgoing waves will be reflected back towards the null point. This means that all the energy in the wave motions is contained within a fixed region.

Craig \& McClymont (\citeyear{CM1993}) investigated the normal mode solutions for both $m=0$ and $m\ne 0$ modes with resistivity included. Again they emphasised that the current builds up as the inverse square of the radial distance from the null point. Craig \& McClymont (\citeyear{CM1993}) also explicitly report on the focusing of the wave (energy) onto the neutral point due to the gradient in Alfv\'en speed (clearly seen in equation \ref{focusss}).

Independently, Hassam (\citeyear{Hassam1992}) also investigated the behaviour of stressed X-points in a cylindrically symmetric geometry. Hassam performed a similar derivation to that of equation (\ref{craig}) and recognised that equation (\ref{f_r})  can be recast as a hypergeometric equation  (e.g. Oberhettinger \citeyear{Oberhettinger}) when $m=0$:
\begin{eqnarray*}
(z-1)\frac{d}{dz}\left( z \frac{d}{dz} f\right) = \left( \frac {\lambda}{2}\right) ^2 f \;\;,\label{hypergeometric}
\end{eqnarray*}
using the transformation $z= -r^2 / \eta \lambda$.  The relaxation time found using this formulation is $ |\ln {\eta} | ^2$ and so is in agreement with that found by Craig \& McClymont (\citeyear{CM1991}; \citeyear{CM1993}).

Craig {\it{et al.}} (\citeyear{Craig2005}) comment that it is surprising that fluid viscosity has been neglected in previous studies, and state that the leading terms in the viscous stress tensor actually dominate the plasma resistivity by many orders of magnitude for typical coronal plasmas (as emphasised  by Hollweg \citeyear{Hollweg1986}). These authors extend  the model of  Craig \& McClymont (\citeyear{CM1993}) to include the effect of (isotropic) scalar fluid viscosity. They find the inclusion of viscosity can have a dramatic effect: for non-reconnective modes ($m>0$), the dissipation rate depends only logarithmically on the magnitude of the dissipative coefficient (thus, for $\eta > \nu$ the problem is dominated by $\ln {\eta}$, and $\nu > \eta$ it is dominated by $\ln {\nu}$). For reconnective disturbances ($m=0$), the oscillatory reconnection is suppressed if  $\nu > \eta$, whereas for  $\eta > \nu$, fast oscillatory reconnection is regained. However, the authors do note that although viscosity can dramatically influence the rate of dissipation in the system, finite resistivity is still required for reconnection to occur.  Craig (\citeyear{Craig2008}) has extended this study to include non-isotropic viscosity (Braginskii \citeyear{Braginskii1965}) and find that the main results of Craig {\it{et al.}} (\citeyear{Craig2005}) are still valid.


These papers have led the way in understanding  MHD motions in the neighbourhood of a 2D X-point. However, the assumed cylindrical symmetry means that the magnetoacoustic disturbances can only propagate either towards or away from the null point,  and attention has been restricted to a circular reflecting boundary, so all outgoing waves are reflected back into the vicinity of the null point. Thus, in a sense there is nowhere else for the wave to propagate except into the null point.

In addition, in all these papers (except Bulanov \& Syrovatskii \citeyear{Bulanov1980} which considered asymptotic limits) the boundary motions move the field lines but do not return them to their original positions.  The Poynting flux induced by this imposed motion provides the energy released in the resultant current sheet.  However, if the boundary motions are simply due to the passing of incoming waves through the boundary, then it is not clear that the null point need collapse and form a current sheet. Furthermore, if this is the case, then it is not clear if the energy in the wave (again due to the Poynting flux through the boundary) will dissipate or simply pass through one of the other boundaries.

Finally, all these papers have assumed that the system is best described in terms of normal modes, where a single normal mode can be thought of as the long time evolution of a system. However, normal mode analysis does not allow us to concentrate on the transient features of the wave propagation.

For these reasons, it is informative to   specifically track the propagation of  boundary-driven (as opposed to internally generated)  asymmetric disturbances into the domain.

\section{Two-dimensional null points in a cartesian geometry}\label{section:cartesian}



{{The first investigation that specifically made the extension away from polar symmetry was performed by Ofman (\citeyear{Ofman1992}) and Ofman {\it{et al.}} (\citeyear{OMS1993}), who  performed nonlinear, resistive 2D MHD calculations of the reconnection in the stressed null point and obtained the  $ | \ln {\eta} | ^2$ scaling law numerically, as well as solving the linear dispersion relation  (equation \ref{f_r}) for all azimuthally nonsymmetric perturbations  ($m > 0$) analytically, in agreement with the results of Craig \& McClymont (\citeyear{CM1991}; \citeyear{CM1993}) and Hassam (\citeyear{Hassam1992}).  However, the authors also stressed the significance of the choice of boundary condition, and the strong influence they have on the permissible solutions. Steinolfson {\it{et al.}} (\citeyear{SOM1995}) investigated the effects of the boundary conditions further by perfoming nonlinear, resistive MHD simulations  to study stressed X-points, contrasting both rigid and open boundary conditions. The authors found that, for rigid boundary conditions, they could recover the results of Craig \& McClymont (\citeyear{CM1991}; \citeyear{CM1993}), but found that for open boundary conditions, the X-point was instead deformed by their perturbation to form a current sheet.}}

{{Shortly thereafter,}}  Hassam \& Lambert (\citeyear{HL1996}) utilised a cartesian geometry to investigate the propagation of the Alfv\'en wave in the neighbourhood of a simple X-point (of the form seen in Figure \ref{figure1}a). Using the same coordinate transformations (equation \ref{BS_transform})  as in Bulanov \& Syrovatskii (\citeyear{Bulanov1980}), Hassam \& Lambert again find that the Alfv\'en wave propagates along magnetic fieldlines, i.e. the fluid elements are confined to the magnetic fieldlines they are generated on. However, Hassam \& Lambert also stress the distinction between two different types of boundary driven Alfv\'en wave motions. Perturbations that initially straddle the separatrices eventually accumulate along the separatrices, whereas disturbances that do not initially straddle the separatrices do not accumulate along the separatrices, but leave the system and are eventually damped by phase-mixing (Heyvaerts \& Priest \citeyear{HP1983}).

Hassam \& Lambert investigated the propagation of the Alfv\'en wave in a square numerical box by driving a harmonic wave train (polarised transverse to the plane of the magnetic field) on both the left and right boundaries, simulating periodic footpoint motion. Their choice to drive wave-trains into the domain from opposite sides of the box  means the forcing is  antisymmetric in the horizontal direction. In addition, line-tied conditions are used on the upper and lower boundaries, i.e. both velocity  and the normal derivative of magnetic field are kept zero. Thus, wave motions are reflected back into the numerical domain, limiting the evolution.



An  investigation of  MHD wave propagation in a $\beta=0$ plasma in the neighbourhood of a 2D null point has been looked at in a series of papers by McLaughlin \& Hood (\citeyear{MH2004}; \citeyear{MH2005}; \citeyear{MH2006a}), where the focus was on  more general disturbances, more general boundary conditions and single wave pulses (rather than harmonic wave trains). By looking at boundary-driven disturbances generated from a single boundary, coupled with non-reflecting boundary conditions, such investigation allow us to  focus on the transient features on the propagation.

To clearly demonstrate the {{results of McLaughlin \& Hood}}, we repeat part of their analysis here:

\subsection{Governing equations and coordinate system of McLaughlin \& Hood}

To study the nature of wave propagation near null points, McLaughlin \& Hood utilised the linearised MHD equations. Here, we use subscripts of $0$ for equilibrium quantities and $1$ for perturbed quantities, such that ${\bf{v}}={\bf{0}}+ {\bf{v}}_1$ (no background flows). The only exception is  ${\bf{B}}={\bf{B}}_0+ {\bf{b}}$, where ${\bf{b}}=(b_x,b_y,b_z)$.

McLaughlin \& Hood  now consider a special coordinate system for $\mathbf{v}_1$:
\begin{eqnarray*}
\mathbf{v}_1 = \frac {{\rm{v}}_\parallel}{|\mathbf{B}_0|} \left(  { \frac {\mathbf{B}_0 } { {|\mathbf{B}_0}|}  }  \right) -       \frac  {{\rm{v}}_\perp}{ {|\mathbf{B}_0|}}  \left( {\frac { \nabla A_0 } { {|\mathbf{B}_0|}} } \right)  + v_y \: {\hat{\bf{y}}}    \;\; ,
\end{eqnarray*}
where $A_0$ is the equilibrium vector potential. The terms in brackets are unit vectors. This splits the velocity into parallel and perpendicular components. This will make our MHD mode detection and interpretation easier. For example, in a low $\beta-$plasma, the slow wave is guided by the magnetic field and has a velocity component that is mainly field-aligned. This makes perfect sense when $\beta \ll 1$ but its usefulness is less clear when $\beta \gg 1$ near the null point (recall the properties detailed in Table 1).



Now consider a change of scale to non-dimensionalise; let ${\rm{\bf{v}}}_1 = \bar{\rm{v}} {\mathbf{v}}_1^*$, ${\rm{v}}_\perp =  \bar{\rm{v}} B {\rm{v}}_\perp^*$, ${\rm{v}}_\parallel =  \bar{\rm{v}} B {\rm{v}}_\parallel^*$,  ${\mathbf{B}}_0 = B {\mathbf{B}}_0^*$, ${\mathbf{b}} = B {\mathbf{b}}^*$, $x = L x^*$, $z=Lz^*$, $p_1 = p_0 p_1^*$, $\nabla = \nabla^* / L$, $t=\bar{t}t^*$, $A_0=B L A_0^*$ and $\eta = \eta_0$, where we let * denote a dimensionless quantity and $\bar{\rm{v}}$, $B$, $L$, $p_0$, $\bar{t}$ and $\eta_0$ are constants with the dimensions of the variable they are scaling. We then set ${B}/{\sqrt{\mu \rho _0 } } =\bar{\rm{v}}$ and $\bar{\rm{v}} =  {L}/{\bar{t}}$. We also set ${\eta_0 \bar{t} } / {L^2} =R_m^{-1}$, where $R_m$ is the magnetic Reynolds number, and set $ {\beta_0} = {2 \mu p_0}/{B^2}$, where $\beta_0$ is the plasma-$\beta$ at a distance unity from the null/origin.

This process non-dimensionalises the linearised MHD equations. For the rest of this section, we drop the star indices; the fact that the variables are now non-dimensionalised is understood.  Thus, the linearised, non-dimensionalised equations are:
\begin{eqnarray}
\rho_0  \frac{\partial }{\partial t }  {{\rm{v}_\perp}}  &=& - { |\mathbf{B}_0 |^2 }   \left[ \left( \nabla  \times  \mathbf{b} \right) \cdot {\hat{\bf{y}}} \right] + {\frac {\beta_0}{2}}  \nabla  A_0  \cdot \nabla  p_1   \nonumber \; \;, \\
\rho_0  \frac{\partial }{\partial t }  {{\rm{v}_\parallel}}   &=& - {\frac {\beta_0}{2}} \left( \mathbf{B}_0  \cdot \nabla  \right) p_1           \nonumber  \;     \;,\\
\rho_0  \frac{\partial {{{v} _y}}}{\partial t } &=& \left( { \mathbf{B}} _0 \cdot \nabla  \right)  b_y \; \;, \nonumber \\
\frac{\partial {b}_x  }{\partial t } &=& \left[ \left(\nabla  {{\rm{v} _\perp}} \times {\hat{\bf{y}}}\right)  \cdot {\hat{\bf{x}}} \right] + \frac{1}{R_m} \nabla^{2} {{b}}_x      \nonumber   \;  \;, \\
\frac{\partial {b} _y }{\partial t } &=&  \left( { \mathbf{B}} _0 \cdot \nabla  \right) v_y  + \frac{1}{R_m} \nabla^{2} {{b}}_y    \nonumber  \;   \;, \\
\frac{\partial {b}_z  }{\partial t } &=& \left[ \left(\nabla  {{\rm{v} _\perp}} \times {\hat{\bf{y}}}\right)  \cdot {\hat{\bf{z}}} \right] + \frac{1}{R_m} \nabla^{2}{{b}}_z      \nonumber  \;   \;, \\
\frac{\partial p_1  }{\partial t } &=& -\gamma \left[ \nabla  \cdot \left( \frac {\mathbf{B}_0  {{\rm{v}_\parallel}} }{   |\mathbf{B}_0 |^2  }\right) - \nabla  \cdot\left( \frac {{{\rm{v}_\perp}}  \nabla  A_0 }{  | \mathbf{B}_0 |^2  }\right) \right]  \label{GOVERNING_ddd}  \; \;.
\end{eqnarray}


Note that the 2D geometry considered by McLaughlin \& Hood is in the  $xz-$plane. As  noted first by Bulanov \&  Syrovatskii (\citeyear{Bulanov1980}), this means that the  Alfv\'en wave (in the  ${\hat{\bf{y}}}-$direction) is decoupled from the magnetoacoustic waves (in the  $xz-$plane).




Finally, we note that equations (\ref{GOVERNING_ddd}) simplify greatly if we neglect pressure perturbations (i.e. assume $\beta=0$).  Under the $\beta=0$ assumption, the plasma pressure plays no part in the dynamics of the system, and so the linearised equation of mass continuity has no influence on the momentum equation and so in effect the plasma is arbitrarily compressible (Craig \& Watson \citeyear{CW1992}) and we assume the background gas density is uniform ($\rho_0$). A spatial variation in $\rho _0$ can cause phase mixing (Heyvaerts \& Priest \citeyear{HP1983}; De Moortel {\it{et al.}} \citeyear{DeMoortel1999}; Hood {\it{et al.}} \citeyear{Hood2002}).

The ideal, $\beta=0$, linearised MHD equations naturally decouple into two equations for the fast magnetoacoustic wave (governed here by  $ {{\rm{v}_\perp}}$)   and for the Alfv\'en wave (governed by $v_y$). The slow wave is absent in the $\beta=0$ limit ($ {{\rm{v}_\parallel}}=0$).

Under these assumptions, the linearised equations for the fast magnetoacoustic wave can be combined to form a single wave equation:
\begin{eqnarray}
  \frac{\partial ^2 }{\partial t^2}  {\rm{v}_\perp}=  v_A^2 \nabla ^2  {\rm{v}_\perp}   = v_A^2 \left( \frac{\partial^2 }{\partial x^2} + \frac{\partial ^2 }{\partial z^2}  \right)  {\rm{v}_\perp}     \; \;,   \label{fast_wave}
\end{eqnarray}
where $v_A \left( x,z \right)= {|{\bf{B}}_0|/ \sqrt{\rho_0}} = \sqrt{ \left(B_x^2+B_z^2 \right)/ \rho_0 }$ is the equilibrium (unperturbed)  Alfv\'{e}n speed.

Similarly, the linearised equations for the Alfv\'en wave can be combined to form a single wave equation:
\begin{eqnarray}
\frac{\partial ^2 }{\partial t^2}  v_y =  \left( { \mathbf{B}}_0 \cdot \nabla \right)^2 v_y = \left(B_x \frac {\partial }{\partial x} +B_z\frac {\partial }{\partial z} \right) ^2 v_y \; . \label{alfven_wave}  
\end{eqnarray}

Wave equations (\ref{fast_wave}) and (\ref{alfven_wave}) are the primary equations governing the behaviour of the linear, $\beta=0$, fast and Alfv\'en waves in an equilibrium magnetic field ${\bf{B}}_0$. These equations form the basis of the investigations carried out by McLaughlin \& Hood (\citeyear{MH2004}; \citeyear{MH2005}; \citeyear{MH2006a}) for various magnetic configurations.


\subsection{Single two-dimensional null point}

McLaughlin \& Hood (\citeyear{MH2004}) investigated the behaviour of the fast and Alfv\'en waves about a simple 2D X-type null point using the following equilibrium magnetic field:
\begin{eqnarray}
{\bf{B}}_0 =\frac{B}{L} \left( x,0,-z \right)   \;\;.\label{simple_X_point}
\end{eqnarray}
This magnetic field represents a $\pi / 4$ rotation  of the magnetic field seen in Figure \ref{figure1}$a$. Note that this particular choice of magnetic field is only valid in the neighbourhood of the null point located at $x=0$, $z=0$.

\begin{figure*}
\hspace{-1.0cm}
\includegraphics[width=6.0in]{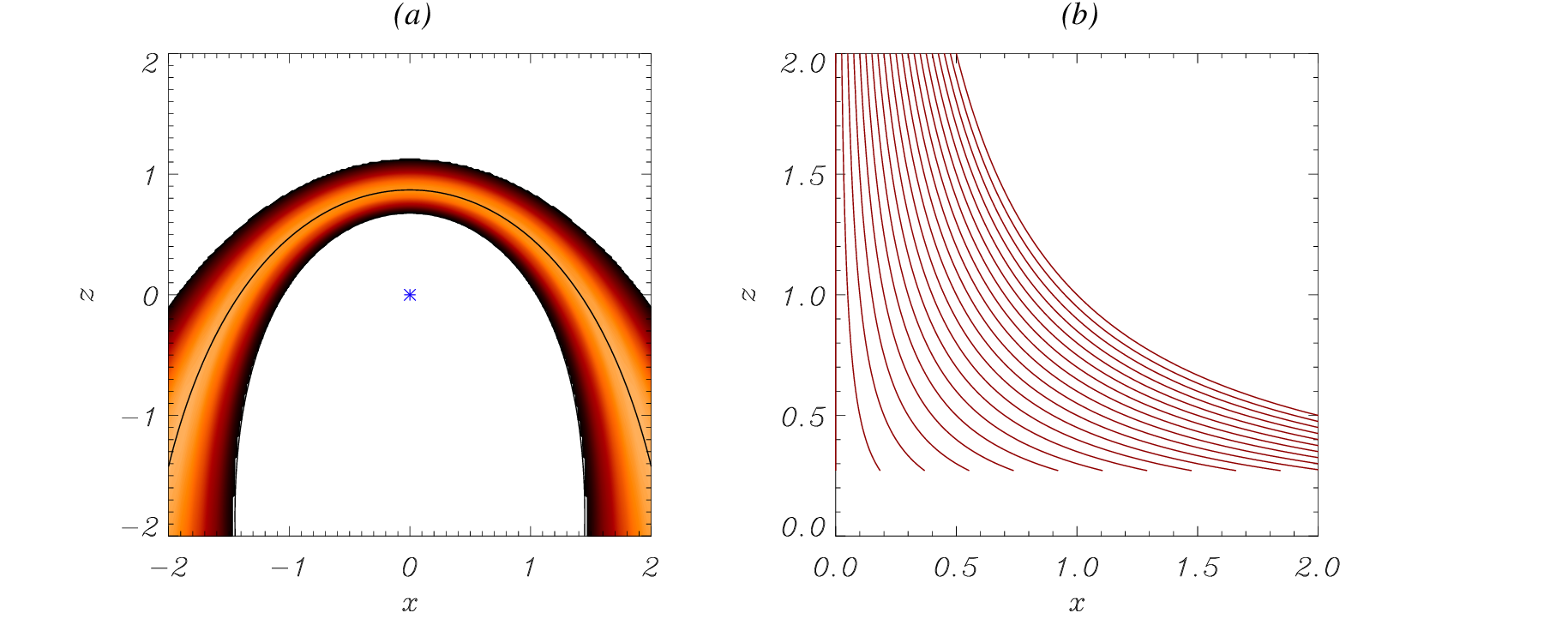}
\caption{$(a)$ Contour of ${\rm{v}}_\perp$ at $t=2$ for a fast wave driven from the upper boundary. The overplotted black lines are the WKB solution, where the three lines represent the leading, middle and trailing edges of the wave pulse. The blue cross denotes the null point (located at the origin). $(b)$ Ray paths of the WKB solution for an Alfv\'en wave driven at the upper boundary,  after a time $t=2$, for starting points of $x=0,  0.025,  ... 0.5$.}
\label{figure5}
\end{figure*}

McLaughlin \& Hood (\citeyear{MH2004})  considered a single wave pulse coming in from the top boundary of the form:
\begin{eqnarray}
   {\rm{v}_\perp}  (x,   z_{\rm{max}}   ) &=& \left\{\begin{array}{cl}
{\sin { \omega t } } & {\mathrm{for} \; \;0 \leq t \leq \frac {\pi}{\omega} } \label{wave_pulse}\\
{0} & { \mathrm{otherwise} }
\end{array} \right. \; , \\
\left. \frac {\partial   {\rm{v}_\perp} } {\partial x } \right| _{x= x_{\rm{min}}} &=& 0 \;\; , \;\; \left. \frac {\partial   {\rm{v}_\perp} } {\partial x }  \right| _{x= x_{\rm{max}}} = 0 \;\; , \;\;\left.  \frac {\partial   {\rm{v}_\perp} } {\partial z }  \right| _{z=    z_{\rm{min}}  }  = 0 \; \;,\nonumber
\end{eqnarray}

where $\omega$ is the frequency. The authors find that the linear fast magnetoacoustic wave travels towards the neighbourhood of the X-point and bends around it. Since the Alfv\'en speed, $v_A(x,z) = {|{\bf{B}}_0|/ \sqrt{\rho_0}} = x^2+z^2$, is spatially varying, the wave travels faster the further it is away from the null point. Thus,  the wave demonstrates {\emph{refraction}} and this can be seen in Figure \ref{figure5}a. A similar refraction phenomenon was found by Nakariakov \& Roberts (\citeyear{Nakariakov1995}). This refraction effect wraps the wave around the null point, and it is this that is the key feature of linear, $\beta=0$ fast wave propagation. This refraction effect is the (non-radial) generalisation of the (purely radial) focusing effect reported by Bulanov \& Syrovatskii (\citeyear{Bulanov1980}) and Craig \& McClymont (\citeyear{CM1991}; \citeyear{CM1993}).


Note, since the Alfv\'en speed drops to zero at the null point, the wave never reaches there, but the length scales (which can be thought of as the distance between the leading and trailing edges of the wave pulse) rapidly decrease, indicating that gradients, and hence the current density, will rapidly increase. McLaughlin  \& Hood (\citeyear{MH2004}) show that for the simple 2D X-point, all gradients increase exponentially as they approach the null point. The rate of this build-up is extremely important, as it implies that resistive dissipation will eventually become important, regardless of the size of $\eta$, and this will convert  the wave energy into (ohmic) heat. In fact, the exponential growth of the current density indicates that the time for magnetic diffusion to become important will depend on $\ln \eta $. 

This is in good agreement with Craig \& McClymont (\citeyear{CM1991}; \citeyear{CM1993}) and  Craig \& Watson (\citeyear{CW1992}) who had previously found that the reconnection rate scales as $|\ln \eta |^2$. This means that wave dissipation will be very efficient, and predicts that null points will be the natural locations of linear fast wave energy deposition and preferential heating.



To confirm their results, McLaughlin  \& Hood (\citeyear{MH2004}) also solved equation  (\ref{fast_wave}) approximately using the WKB approximation (e.g. Murray \citeyear{Murray1927}; Sneddon \citeyear{Sneddon1957}). The WKB solution of McLaughlin \& Hood assumes $\omega \gg 1$  and is obtained using the method of characteristics (e.g. Bender \& Orszag \citeyear{BO1978}) and is in excellent agreement with the original numerical solution. This can be seen from the overplot in Figure \ref{figure5}a.

By  considering equation (\ref{alfven_wave}), McLaughlin  \& Hood (\citeyear{MH2004}) find that the linear Alfv\'en wave propagates down from the upper boundary and begins to spread out, following the field lines, in agreement with Bulanov \& Syrovatskii (\citeyear{Bulanov1980}). The wave is confined to the fieldlines it is excited on. As the wave approaches the separatrix (defined by the $x-$axis), the pulse thins but keeps its original amplitude. The wave eventually accumulates along the separatrices. As for the fast wave, we have decreasing length scales, and for this choice of set-up, $j_x$ grows exponentially in time. Hence, the authors find that the Alfv\'en wave causes current density to accumulate along the separatrices (in agreement with  Hassam \& Lambert \citeyear{HL1996}). Hence, all the {\emph{Alfv\'en wave energy will be dissipated along the separatrices}} and these will be the locations for preferential heating.

A WKB solution was also obtained for the Alfv\'en wave and this was in excellent agreement with the numerical results. In Figure \ref{figure5}b, we can see the evolution of fluid elements that begin at points $x=0,  0.025 ,  ... 0.5$, which clearly demonstrates  how the fluid elements simply travel along the fieldlines they start on. In addition, note that at $t=2$ the fluid elements have all travelled different distances along their respective fieldlines, but the wave remains planar. McLaughlin \& Hood explain this and  show that the leading edge has in fact reached $z=z_{\rm{max}}e^{-t} = 2e^{-2}=0.27$.

Note that the behaviour of the Alfv\'en wave is different to that of the fast wave in the sense that the two wave types deposit all their wave energy at different areas (along separatrices as opposed to at the null point). However, the phenomenon of depositing wave energy in a specific area is common to both. In addition, this work is in good agreement with previous work in cylindrically symmetric geometries (\S\ref{section:polar}). Thus, some general and robust properties of the fast and Alfv\'en wave are becoming apparent.



\subsection{Pair of two-dimensional null points}


McLaughlin  \& Hood (\citeyear{MH2005}) repeated the aboce investigation for a magnetic field containing two null points. Such a configuration is particularly relevant since it can be argued that null points appear in pairs. For example,  a double null point may arise as a local  bifurcation of a single 2D null point (see e.g. Galsgaard {\it{et al.}} \citeyear{KRRN1996}; Brown \& Priest \citeyear{BP1998}).



Two magnetic configurations were considered; one containing a separator and one that does not.  The magnetic configuration with a separator is taken as:
\begin{equation}
{\bf{B}}_0 = \frac{B}{L^2} \left({x^2-z^2-\lambda^2}, 0, {-2xz}\right)\;\;,\label{eq:left}
\end{equation}
where $2\lambda$ is the distance between the null points. Note that this introduces a characteristic length scale into the system, whereas previously the single X-point had no charactistic length scale. The other equilibrium magnetic field considered takes the form:
\begin{equation}
{\bf{B}}_0 = \frac{B}{L^2} \left({2xz}, 0, {x^2-z^2-\lambda^2}\right)\;\;.\label{eq:right}
\end{equation}
These equilibrium magnetic field configurations can be seen in Figures \ref{figure2}b and \ref{figure2}c. The authors considered the $\beta=0$ linearised MHD equations and solved equations  (\ref{fast_wave}) and (\ref{alfven_wave}) for these two magnetic equilibria. 


McLaughlin \& Hood (\citeyear{MH2005}) find that for the linear fast magnetoacoustic  wave approaching the two null points from above, the wave travels down towards the null points and begins to refract around them both. The wave the \lq{breaks}\rq{ } into two along the line $x=0$ (due to symmetry), with each half of the wave going to its closest null point. Each part of the wave then continues to wrap around its respective null point repeatedly, eventually accumulating at that specific null point.


In the case of the fast wave pulse travelling in from the side boundary, we see a similar effect (i.e. a refraction effect, wave breakage and accumulation at the nulls), but in this case the wave is \emph{not} equally shared between the null points. For example,  for a fast wave travelling in from the left boundary, initially the pulse thins, begins to feel the effect of the left-hand-side null point and begins to refract around this null. As the ends of the wave wrap around behind the left null point, they then become influenced by the right-hand-side null point. These arms of the wave then proceed to wrap around the right null point, flattening the wave.  Furthermore, the two parts of the wave now travelling through the area between the null points have non-zero  Alfv\'{e}n speed, and so can propagate through this area. These parts of the wave break along $x=0$ and then proceed to wrap around the null point closest to them.

As before, it is clear the refraction effect focuses all the energy of the incident wave towards the null points, but McLaughlin \& Hood (\citeyear{MH2005}) find that the  angle that the fast wave approaches the null points from will determine what proportion of wave energy ends up at each null point (i.e. where the wave \lq{breaks}\rq{ }). In the case of the fast wave, all the wave energy is accumulating at the null points and since we have a changing perturbed magnetic field with increasing gradients,   this is where current density will accumulate. Thus,  fast wave heating will naturally occur at both null points.



For the Alfv\'en wave, the results show that the wave propagates along the field lines,  thins but keeps its original amplitude,  and eventually accumulates along the separatrices (again, in agreement with previous studies).


Thus, McLaughlin \& Hood (\citeyear{MH2005})  find that the key results from McLaughlin \& Hood (\citeyear{MH2004}) carry over from a single 2D null configuration to that of a pair of 2D null points.

\subsection{Single null point configuration created by  two magnetic dipoles}

McLaughlin \& Hood (\citeyear{MH2006a}) again investigate the behaviour of the fast and Alfv\'en wave in an ideal,  $\beta=0$ plasma, but now address a key problem with the first two papers: that the simple null points considered (i.e. equations \ref{simple_X_point}, \ref{eq:left}, \ref{eq:right}) are only valid locally, because as $x$ and $z$ get very large, ${\bf{B}}_0$ also gets unphysically large.

To address this issue, McLaughlin \& Hood (\citeyear{MH2006a}) investigate the behaviour of MHD waves near an equilibrium magnetic field created by two dipoles:
\begin{eqnarray}
B_x &=&\;\;\; BL^2 \left( \frac{ - \left( x+\lambda\right)^2 + z^2 }{ \left[\left( x+\lambda\right)^2 +z^2\right]^2} +   \frac{ - \left( x-\lambda\right)^2 + z^2 }{ \left[\left( x-\lambda\right)^2 +z^2\right]^2}  \right)\; ,\nonumber\\
B_z&=&-BL^2 \left(\frac { 2\left( x+ \lambda \right)z}{ \left[\left( x+ \lambda \right)^2 +z^2\right]^2} -  \frac { 2\left( x- \lambda \right)z}{ \left[\left( x- \lambda  \right)^2 +z^2\right]^2} \right)\label{twodipoles}\;\;,
\end{eqnarray}
where $2\lambda$ is the separation of the dipoles. This magnetic field (with $2\lambda=1$) can be seen in Figure \ref{figure1}c. It comprises of four separatrices and an X-point located at $(x,z)=(0,\lambda)$. Note that as $x$ or $z$ gets very large, the field strength becomes small. Hence,  this is a more physical field than those previously investigated.


McLaughlin \& Hood (\citeyear{MH2006a}) drive a fast-wave planar pulse on their lower boundary (along $z=0$). As the wave propagates upwards, the planar wave is distorted by the   two regions of high Alfv\'en speed and the wave forms two peaks with maxima located over the loci of the magnetic field. Close to the null point, the fast wave begins to refract around the null point. Meanwhile, the rest of the wave (referred to as the \emph{wings}) continue to propagate upwards and spread  out (since the fast wave propagates roughly isotropically). The wave is stretched between its two goals (part wrapping around the null and part travelling away from the magnetic skeleton) and this leads to the wave splitting; near the regions of high Alfv\'en speed the localised high speed thins the wave and forces the split. Thus, part of the (now split) wave spirals into the null and the other part propagates away from the magnetic skeleton.  Finally,  a WKB solution demonstrates that there is a critical radius of influence within which a fast wave will be captured by the null point and that, for the parameters considered by McLaughlin \& Hood,   $40\%$ of the  wave is trapped by the null. This makes intuitive sense: if the fast wave is too far away from the magnetic null, it will not feel its effect.


McLaughlin \& Hood (\citeyear{MH2006a}) also looked at the behaviour of the Alfv\'en wave and found that, as before, the propagation follows the magnetic fieldlines but that now only  part of the wave accumulates along the separatrices and the other part of the wave  appears to propagate away from the magnetic skeleton. However, the Alfv\'en wave is  actually just following the fieldlines (and these are spreading out) recovering the key results of  Hassam \& Lambert (\citeyear{HL1996}), i.e. there is a key difference between the evolution of perturbations that initially straddle the separatrices and those that do not.


\subsection{MHD wave behaviour  at $\beta \neq 0$ null points}\label{section:mode conversion}

The key results from McLaughlin \& Hood (\citeyear{MH2004}; \citeyear{MH2005}; \citeyear{MH2006a}) demonstrate that the behaviour of the fast wave in a  $\beta=0$ plasma is entirely dominated by the Alfv\'en-speed profile, and since the magnetic field drops to zero at the X-point, the wave will never reach the actual null. The  next step is to extend the model to include plasma pressure ($\beta \neq 0$ plasma).  The most obvious effect is the introduction of {\emph{slow magnetoacoustic waves}}. The fast wave can now also pass through the null point (there is a non-zero sound speed there) and thus perhaps take energy away from that area. Such a model could also involve mode coupling in areas where the sound speed and Alfv\'en speed become comparable in magnitude.

Such an investigation has been carried out into the behaviour of magnetoacoustic waves by McLaughlin \& Hood (\citeyear{MH2006b}) for a single 2D null point (i.e. equilibrium magnetic field given by equation \ref{simple_X_point}). Note that the plasma pressure plays no role in the  propagation of the linear Alfv\'en wave, and so the description by McLaughlin \& Hood (\citeyear{MH2004})  remains valid.


In most parts of the corona, the plasma-$\beta$  (equation \ref{plasmabetaequation}) is much less than unity and hence the pressure gradients in the plasma can be neglected. However, near null points the magnetic field vanishes  and so the plasma-$\beta$ can become very large. Thus, understanding the changing plasma-$\beta$ is of key importance here. Considering equilibrium quantities;
\begin{eqnarray*}
 \beta=\frac{2 \mu p_0L^2}{B^2 \left( x^2+z^2\right)}\;\; \Rightarrow\; \beta=\frac{\beta_0}{x^2+z^2} =\frac{\beta_0}{r^2} \;\;,
\end{eqnarray*}
where $r^2=x^2+z^2$ and $\beta_0={2 \mu p_0L^2}/{B^2}$.

Thus, the plasma-$\beta$ varies through the whole region, since magnetic  field is varying everywhere throughout our model. In fact, the plasma$-\beta$ is infinite at the null point. In particular, we note that outside a radius of unity we have a low-$\beta$ environment and inside we have a high$-\beta$ environment. This will have important consequences as fast and slow waves have differing properties depending upon their environment (see Table 1).

There is also coupling between the perpendicular and parallel velocity components (when $\beta \neq 0$) and this coupling is most effective where the sound speed and the Alfv\'en velocity are comparable in magnitude.  Bogdan {\it{et al.}} (\citeyear{Bogdan2003}) call this zone the {\emph{magnetic canopy}} or the  $\beta \approx 1$ layer. Note that here, we use the terminology $\beta$ for the (true, varying) plasma-$\beta$ and $\beta_0$ for the (constant) value of the  plasma-$\beta$ at a radius of unity.

The  $\beta=1$ layer  occurs at radius $r = \sqrt{\beta_0}$.  However,  it is not the $\beta=1$ layer that is most important in understanding this system, but the layer where the sound speed is equal to the Alfv\'en speed, i.e.  $c_s=v_A$.  Recalling that $c_s= \sqrt{{\gamma  \beta_0} / 2} v_A $, this means that the  $c_s=v_A$ layer  (or alternatively the $\beta={2}/{\gamma}$ layer) occurs at a radius  $r= \sqrt{{\gamma} \beta_0 / 2}$.  Of course, the difference between the $\beta=1$ layer at  $r = \sqrt{\beta_0}$  and the $c_s=v_A$ layer at $r= \sqrt{{\gamma} \beta_0 / 2}$  is very small, and hence it is easier to refer to  the $\beta \approx 1$ layer.

{{Finally, we note that the basic fast wave speed in this $\beta \neq 0$ system is:
\begin{eqnarray*}
c^2_{{\rm{fast}}} = {\frac{1}{2}}\left(v_A^2+ c_s^2\right)   + {\frac{1}{2}} \sqrt{ \left(v_A^2+ c_s^2\right)^2 - 4 \:v_A^2\: c_s^2\: {\cos^2 {\theta}} }
\end{eqnarray*}
which for perpendicular propagation reduces to
\begin{eqnarray}
c^2_{{\rm{fast}}} = v_A^2+ c_s^2  = \frac{\gamma}{2} \beta_0 + x^2 + z^2 \;\;.\label{herehere}
\end{eqnarray}

}}

McLaughlin \& Hood (\citeyear{MH2006b}) solve the linearised MHD equations  (\ref{GOVERNING_ddd}) with $\beta_0 \neq 0$. Using identical boundary conditions to McLaughlin \& Hood (\citeyear{MH2004}), i.e. equation \ref{wave_pulse},  a wave pulse is driven in the perpendicular velocity component on the upper boundary, which corresponds to driving a low-$\beta$ fast wave.  It is found that  the low-$\beta$ fast wave propagates into the neighbourhood of the null point and begins to refract around it. However, as the wave crosses the $c_s=v_A$ layer, the low-$\beta$ fast wave transforms into a high-$\beta$ fast wave (due to the change in environment) and also generates a high-$\beta$  slow wave. The fraction of the incident wave converted into slow wave is found to be proportional to $\beta_0$. The magnetoacoustic propagation now proceeds in three ways:
\begin{itemize}
\item{Firstly, the generated slow wave spreads out along the fieldlines, eventually accumulating along the separatrices.}
\item{Secondly, the remaining part of the fast wave inside the $c_s=v_A$ layer continues to refract and some of it now {\emph{passes across the null}}. We identify this wave as a high-$\beta$ fast wave. The high-$\beta$ fast wave can pass through the null point because, although $v_A (0,0) =0$, there is now a non-zero sound speed there (clearly seen in equation \ref{herehere}). After it has crossed the null, the high-$\beta$ fast wave continues to propagate downwards  and crosses  the $c_s=v_A$ layer for a second time. As it emerges, the wave now becomes a low-$\beta$ fast wave and  spreads out isotropically.}
\item{Finally, the (low-$\beta$) fast wave located away from the null and away from the $c_s=v_A$ layer  (again referred to as the \lq{wings}\rq{ } of the low-$\beta$ wave) are not affected by the non-zero sound speed (as $v_A\gg c_s$) and so here the refraction effect dominates. In fact, as these wings wrap around below the null point, they encounter the high-$\beta$ fast wave as it is emerging from the $c_s=v_A$ layer. This results in a complicated interference pattern, but it appears that the two waves pass through  each other (due to the linear nature of the system).}
\end{itemize}


It is clear that there are two competing phenomena: $(a)$ the {\emph{refraction effect}} due to the varying Alfv\'en speed and $(b)$ a non-zero sound speed at the null which allows the fast wave to pass through it. It is the value of  $\beta_0$ that dictates which effect dominates.  Thus, two extremes can occur. The first occurs when $\beta_0 \to 0$, in which case the refraction effect dominates  and we recover the results of  McLaughlin \& Hood (\citeyear{MH2004}). The second occurs when $\beta_0 \to \infty$ and the system becomes hydrodynamic. In this case, the fast wave reduces to an acoustic wave and so completely passes through the null (effectively, it does not even see the magnetic field, since $v_A \ll c_s$). Thus, it is possible to understand the whole spectrum of values of the parameter $\beta_0$.

McLaughlin \& Hood (\citeyear{MH2006b}) also noted a tendency for their system to develop several lobe-like structures in various variables. This occurs because the choice of magnetic equilibrium {\emph{naturally}} leads to a ${\sin { 2 \theta}}$ and ${\sin { 4 \theta}}$ dependence in ${\rm{v}}_\parallel$ when ${\rm{v}}_\perp$ is driven (this was demonstrated in polar coordinates, using equilibrium magnetic field \ref{simple_X_point_polar}). Thus, with this choice of null point, if we drive any of the velocity variables then the system will naturally develop a $\theta-$dependence.

Finally, the authors compared their results with an analytical WKB approximation. This resulted in two separate wave descriptions, corresponding to the fast or slow wave. However, the WKB solution could only reproduce the propagation of the fast wave (in both its low and high-$\beta$ environments) or the slow wave  but not both together. Instead, the WKB solution broke down at the conversion layer (at $c_s=v_A$) where the approximation becomes degenerate.  Thus, the WKB solution (in the form presented by McLaughlin \& Hood) cannot be used to investigate mode conversion. Instead, addition terms are needed in the approximation (the authors  only consider the first-order terms in the WKB approximation). Alternatively, this degeneracy can be overcome by using the method developed by Cairns \& Lashmore-Davies (\citeyear{Cairns}) to match WKB solutions across the mode conversion layer. This has been done in 1D (McDougall \& Hood \citeyear{Dee}) but a 2D investigation has yet to be completed.


\subsection{Nonlinear simulations in the neighbourhood of a 2D X-type null point}\label{section:nonlinear}

McLaughlin \& Hood (\citeyear{MH2004}; \citeyear{MH2005}; \citeyear{MH2006a}; \citeyear{MH2006b}) investigated the behaviour of the linear fast and slow magnetoacoustic waves and Alfv\'en waves in the neighbourhood of a variety of 2D null points using equations (\ref{GOVERNING_ddd}).  However, the validity of the linearisation is questionable once the perturbed velocity becomes comparable to the magnitude of the local Alfv\'en speed. McLaughlin {\it{et al.}} (\citeyear{MDHB2009}) extend the models of McLaughlin \& Hood (\citeyear{MH2004}) and Craig \& McClymont(\citeyear{CM1991}) to include nonlinear effects in a $\beta \neq 0$ plasma, and consider  the behaviour of the nonlinear fast wave. The authors solved the nonlinear, compressible, resistive MHD equations (equations \ref{MHDequations})  using a Lagrangian-remap, shock-capturing code ({\emph{LARE2D}}, Arber {\it{et al.}} \citeyear{LARE}), for a simple 2D X-point configuration:
\begin{eqnarray*}
{\bf{B}}_0 =\frac{B}{L} \left(y,x,0 \right)   \;\;,
\end{eqnarray*}
which corresponds to the magnetic field seen in Figure \ref{figure1}a.{{ Note that Ofman (\citeyear{Ofman1992}), Ofman {\it{et al.}} (\citeyear{OMS1993}) and  Steinolfson {\it{et al.}} (\citeyear{SOM1995}) had previously performed nonlinear 2D calculations of stressed X-points.}}

McLaughlin \& Hood (\citeyear{MH2004}) and previous results from investigations conducted in a cylindrical geometry ($\S\ref{section:polar}$) clearly demonstrate  that the Alfv\'en speed (${\rm{v}}_A^2=B_x^2+B_y^2=x^2+y^2=r^2$) plays a vital role. Hence, it is natural to consider either a polar coordinate system or to drive a circular pulse. In addition, as commented by  McClements {\it{et al.}} (\citeyear{McClements}),  a disturbance initially consisting of a plane wave is refracted as it approaches the null in such a way that it becomes more azimuthally-symmetric. Thus, as a first step in investigating the nonlinear regime, it is appropriate to consider an azimuthally-symmetric initial condition in velocity  of the form:
\begin{eqnarray}
{\rm{v}}_\perp \left(x,y,t=0 \right)  &=& 2C \sin \left[ \pi \left(r - 4.5 \right) \right]{\rm{\;\; for\;\;\;}}4.5\le r \le 5.5 \label{ICs}\;,\\
{\rm{v}}_\parallel \left(x,y,t=0 \right)  &=& 0    \;\;  ,  \nonumber
\end{eqnarray}
which corresponds to a circular, sinusoidal wave pulse in ${\rm{v}}_\perp$,  where $2C$ is the initial amplitude. When the simulation begins, this initial pulse  naturally splits into two waves, each of amplitude $C$: an outgoing wave and an incoming wave. The authors  focus on the incoming wave, i.e. the wave travelling towards the null point, and set $C=1$. 



\begin{figure}
\hspace{-1.0cm}
\includegraphics[scale=0.38]{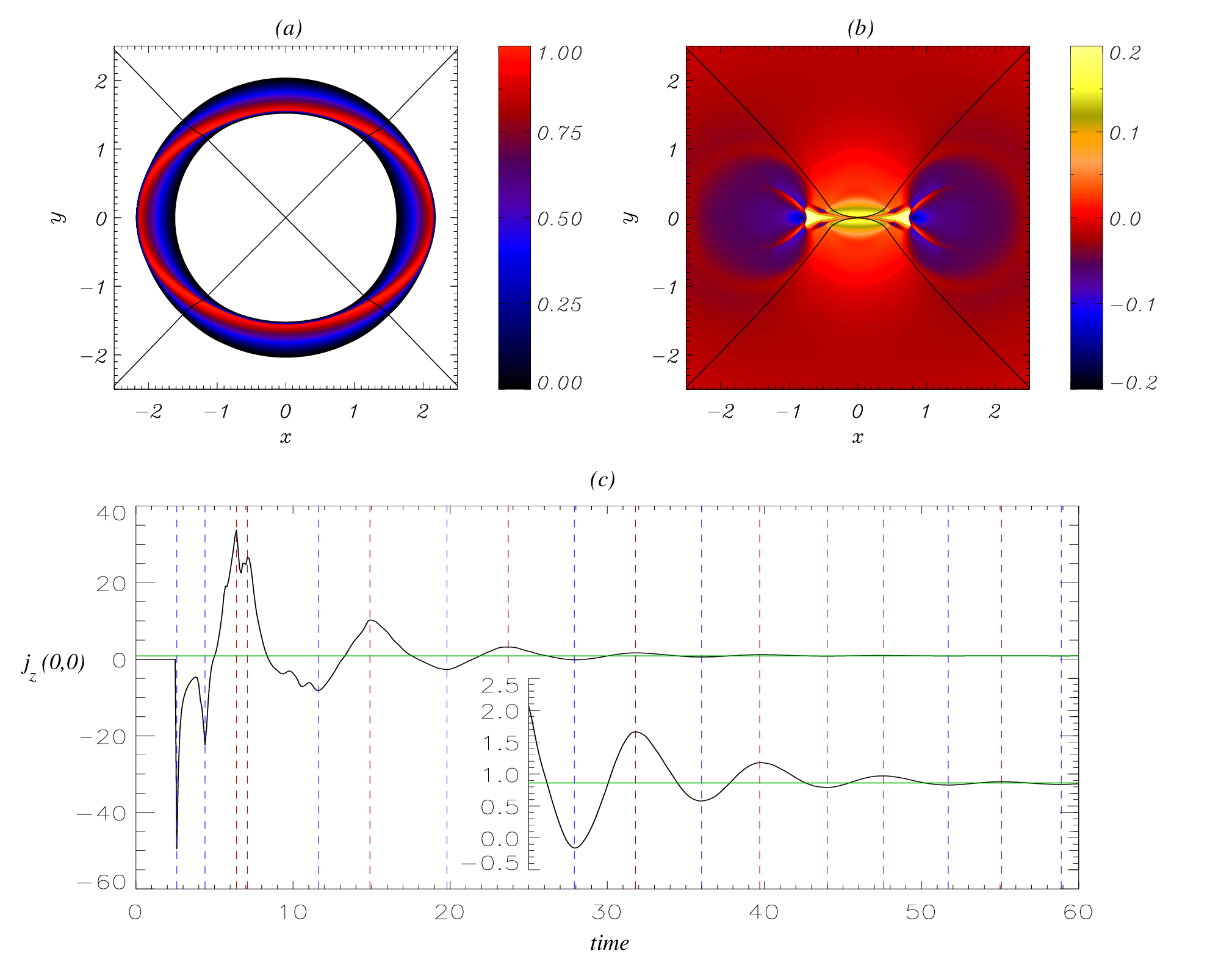}
\caption{Contours of ${\rm{v}}_\perp$ for a fast wave pulse initially located at a radius $r=5$ and its resultant propagation at $(a)$ time $t=1$ and $(b)$ time $t=2.6$. The black lines denote the (changing) separatrices and the null point is located at their intersection (origin). Note in $(b)$ that the separatrices have been deformed and now form a \lq{cusp-like}\rq{} magnetic field structure. The amplitude of ${\rm{v}}_\perp$ varies substantially throughout the evolution, and hence each subfigure is assigned its own colour bar. $(c)$  Plot of time evolution of $j_z (0,0)$ for $0 \le t \le 60$ . Insert shows time evolution of $j_z (0,0)$ for $25 \le t \le 60$ (i.e. same horizontal axis, different  vertical axis). Dashed lines indicate maxima (red) and minima (blue). Green line shows limiting value of  $j_z(0,0)=0.8615$.}
\label{figure6}
\end{figure}

The authors find that at early times  the incoming wave (identified as a linear fast wave) propagates across the magnetic fieldlines  and that the initial pulse profile (an annulus)  contracts as the wave approaches the null point. This is the same refraction behaviour that has previously been reported. The authors also note that the incoming wave pulse develops an asymmetry, where in the $y-$ / $x-$ direction the wave peak / trailing footpoint is catching up with the leading footpoint / wave peak, and eventually forms discontinuities. This can be seen in Figure \ref{figure6}a. The asymmetry develops directly due to the choice of a velocity initial condition. In the nonlinear regime, specifying an initial condition in velocity also prescribes a background velocity profile. Thus, the initial condition (equation \ref{ICs}) appears to excite the $m=0$ mode, but this actually corresponds to the $m=2$ mode in cartesian components.

At later times, these discontinuities develop into fast oblique magnetic shock waves, leading to local heating of the plasma. In addition, the shocks above and below $y=0$ began to overlap, forming a triangular \lq{cusp}\rq{} (called the shock-cusp) and this leads to the development of hot jets, which again heat the local plasma and significantly bent the local magnetic fieldlines. The hot jets (which fit the description of Forbes \citeyear{Terry}) set up slow oblique magnetic  shock waves emanating from the shock-cusp.  {{In addition, there is evidence of slow shocks along the sides of the jet upstream of the tip and we see kinks in the fieldlines at the tip of the jet, indicative of a fast shock.  Thus, the jet heating itself is accomplished by a combination of slow and fast shocks. It is interesting to note that the jet has a bimodal structure consisting of a hot, narrow jet incased within a broader, lower temperature jet, which is a feature that is not predicted by steady-state reconnection theory.  }}

Eventually, the fast shocks reach the null point and the shocks have deformed the magnetic field such that the separatrices now touch one another rather than intersecting at a non-zero angle (called \lq{cusp-like}\rq{} by Priest \& Cowley \citeyear{PC1975}). This can be seen in Figure \ref{figure6}b. {However, the separatrices continue to evolve and so this field structure is not sustained for any length of time.} The  (deformed) null point itself continues to collapse and forms a horizontal current sheet (again, such behaviour was not seen in the linear systems).

Subsequently, the system evolves as follows: the hot jets to the left and right of the null continue to heat the plasma, which in turn expands. This expansion squashes and shortens the horizontal current sheet, forcing the separatrices apart. The (squashed) horizontal current sheet then returns to a  \lq{cusp-like}\rq{} null point which, due to the continuing expansion from the heated plasma, in turn forms a vertical current sheet. The evolution then proceeds through a series of horizontal and vertical current sheets and displays oscillatory behaviour, in a similar manner to the linear results of Craig \& McClymont (\citeyear{CM1991}). The oscillatory nature of the system can be clearly seen from the time evolution of  $j_z(0,0)$ shown in Figure \ref{figure6}c. The red / blue lines indicate  maxima / minima in the system and the green  line shows $j_z(0,0)=0.8615$, which is the limiting value of the oscillation.

It is interesting to note that the final state is non-potential, though still in force-balance, since there is a finite amount of current left in the system. This is because the plasma pressure  in the final state is greater to the left and right of the null point than that above and below, due to the asymmetric heating from the hot jets. Of course, the system will eventually return to a potential state due to diffusion, but this will occur on a far greater timescale than that considered by the simulation ($t_{\rm{diffusion}} \sim R_m=10^4$).

McLaughlin {\it{et al}} (\citeyear{MDHB2009}) provide two pieces of evidence for reconnection in their system. Qualitatively, they observe changes in fieldline connectivity and quantitatively they look at the evolution of the vector potential at the null point. Since they have both oscillatory behaviour {\emph{and}} evidence for reconnection, they conclude that the system displays {\emph{oscillatory reconnection}} (as detailed by Craig \& McClymont \citeyear{CM1991}).

The authors then extended the study to look at the effect of changing the amplitude, $C$, of their initial condition (equation \ref{ICs}).  They conclude that a larger initial amplitude results in a larger amount of current being left in the system at the end of the simulation, i.e. in the non-potential final state.



Thus, it is clear that the nonlinear behaviour is completely different to that of the linear regime. For example, current density now accumulates at many locations, such as along horizontal or vertical current sheets, along slow oblique magnetic shocks and at the location of shock-cusps. This was not the case in the linear regime, where all the current density accumulated at the null point exponentially in time.

The work of McLaughlin {\it{et al.}} (\citeyear{MDHB2009}) provides a link  between two traditionally separate areas of solar physics: MHD wave theory and reconnection, and is the first demonstration of reconnection {\emph{naturally driven}} by MHD wave propagation.


\subsection{Quasi-Periodic Pulsations}\label{QPPs_section}

Quasi-Periodic Pulsations (QPPs) have been observed in radio, optical and X-ray emission of solar flares (e.g. Inglis {\it{et al.}} \citeyear{INM2008}; {Inglis} \& {Nakariakov} \citeyear{IN2009}; {Nakariakov} \& {Melnikov} \citeyear{NM2009}) and stellar flares (e.g. Mathioudakis {\it{et al.}} \citeyear{Mihalis2003}; \citeyear{Mihalis2006}; Mitra-Kraev {\it{et al.}} \citeyear{U2005}). QPPs are observed in emission intensity as oscillations with a characteristic period (from a few seconds to several minutes) and   occur sometimes stably for several minutes and at other times in short bursts.  Several mechanisms have been proposed for the explanation of QPPs, either linking the observations to MHD oscillations (see Nakariakov \citeyear{Nakariakov2007} for a recent review) or being associated with periodic regimes of magnetic reconnection (e.g. Kliem {\it{et al.}} \citeyear{Kliem2000}; Ofman \& Sui \citeyear{OS2006}).

Fast magnetoacoustic wave behaviour in the neighbourhood of magnetic null points provides two alternative mechanisms. Firstly, Nakariakov  {\it{et al.}} (\citeyear{Nakariakov2006}) extend  the model of McLaughlin \& Hood (\citeyear{MH2004}) to include a harmonic driver. As before, the refraction effect wraps the fast wave around the null point, and large gradients and currents develop. By utilising a harmonic driver, the resultant current growth and accumulation is itself periodic, i.e. the periodicity of the incoming fast waves is efficiently transmitted into the periodic modulation of the current density. In addition, Nakariakov  {\it{et al.}} provide an explanation of where such an incoming fast wave could originate:  fast waves can leak from an oscillating loop situated near a null point but magnetically disconnected from it. Interestingly, this would imply that the period of the observed QPPs is linked to the properties of the neighbouring, oscillating loop.


An alternative explanation may be provided by the work of McLaughlin {\it{et al.}} (\citeyear{MDHB2009}) in which oscillatory reconnection is naturally driven by an incoming fast magnetoacoustic wave. Here, the oscillatory behaviour is due to a cycle of horizontal and vertical current sheets (i.e. a different physical mechanism to that of Nakariakov  {\it{et al.}} \citeyear{Nakariakov2006}). However, further work is needed to develop and generalise the model into a diagnostic tool, and such developments should, for example, consider the consequences of different initial conditions such, as utilising a localised increase in pressure or internal energy in a $\beta\neq 0$ plasma, as well as conduct a detailed parametric study.


\subsection{Phase-Mixing}\label{PM_section}

Finally, we consider a novel piece of work by Fruit \& Craig (\citeyear{FC2006}) whereby the phenomenon of phase-mixing (Heyvaerts \& Priest \citeyear{HP1983}) is invoked in  the viscous and resistive dissipation of standing Alfv\'en waves within a line-tied X-point geometry. Here, the authors consider magnetic fieldlines anchored into a rigid, reflective boundary and excite the system with an initial velocity profile. During the simulation, each fieldline, being rigidly tied at the end, oscillates back and forth, with the oscillation frequency depending upon the length and magnetic field strength of that fieldline. The phase difference between neighbouring lines increases as time evolves, leading to growing cross-gradients, and eventually leading to strong visco-resistive damping.  Thus, the authors conclude that phase-mixing can provide an efficient mechanism for energy dissipation of standing Alfv\'en waves  in the vicinity of 2D null points.

Craig \& Litvinenko (\citeyear{CL2007}) extend the model of  Fruit \& Craig (\citeyear{FC2006}) to include anisotropic (Braginskii bulk) viscosity, and find that the main results are still valid.


\section{Weak guide-field}\label{guide_field}



An obvious and essential extension of the 2D work described so far  is to extend the models to full 3D simulations. However, two extension are possible that could provide useful signposts before the move to 3D. Firstly, it is possible to  extend the model to 2.5D with the addition of a third spatial coordinate, by taking into account an extra Fourier component of the form $e^{imy}$, where $m$ is the azimuthal mode number. Secondly, we can consider the addition of a small axial magnetic field perpendicular to the plane of the magnetic X-point, i.e. a longitudinal guide-field. Note that with the addition of a longitudinal guide-field we are no longer  considering null points: even though the X-point geometry remains, ${\bf{B}} \neq 0$ at the X-point (due to the cross field) and we actually have an X-line. However, since the two systems are so closely related we will still provide a literature review of the relevant papers.



Extending the model to 2.5D and 3D will lead to coupling of all the wave modes, and thus will most likely result in energy accumulating at \emph{both} the separatrices and the null points.

The first work in this area was done by Bulanov {\it{et al.}} (\citeyear{Bulanov1992}). Bulanov {\it{et al.}} considered an X-point magnetic field configuration with a longitudinal (along the X-line) magnetic field $B_\parallel$. The authors write down, for the first time, the governing linearised equations in the ideal $\beta=0$ regime, and show that the fast magnetoacoustic wave and Alfv\'en wave are linearly coupled by the gradients in the field. Furthermore, the authors show that magnetoacoustic perturbations can transform into Alfv\'en waves and vice versa, thus leading to current density accumulation at either the null point or along the separatrices. However, the rate or efficiency of this process is not explored in this analytical work. In the limit of $B_\parallel \to 0$, the two modes are decoupled and the results of 2D work are recovered. 

McClements {\it{et al.}} (\citeyear{McClements2006}) also investigated the coupling of Alfv\'en waves and fast waves in the vicinity of a magnetic X-point with a weak longitudinal guide field present ($B_\parallel \ll B_\perp$), and extended the model of  Bulanov {\it{et al.}} (\citeyear{Bulanov1992}) to include resistive effects. These authors solve the initial value problem for a fast wave being driven by a harmonic Alfv\'en wave train and find that energy is channelled into the fast wave, and that large gradients start to build-up near the X-point, due to the Alfv\'en speed profile. The results indicate that a significant fraction of the Alfv\'en wave energy is converted into fast wave energy. Ben Ayed {\it{et al.}} (\citeyear{BenAyed2009}) extend the work of  McClements {\it{et al.}} (\citeyear{McClements2006}) to include a strong guide-field ($B_\parallel \gg B_\perp$). Again, the authors found that the Alfv\'en wave is coupled into the fast mode, with the coupling strongest  on the separatrices and far from the X-line.

The three works (Bulanov {\it{et al.}} \citeyear{Bulanov1992}; McClements {\it{et al.}} \citeyear{McClements2006}; Ben Ayed {\it{et al.}} \citeyear{BenAyed2009}) all assume linear,  $\beta=0$ plasma, and  McClements {\it{et al.}} (\citeyear{McClements2006}) and  Ben Ayed {\it{et al.}} (\citeyear{BenAyed2009}) concentrate on driving linear Alfv\'en wave trains and observing the coupling to the fast mode.

Conversely, Landi {\it{et al.}} (\citeyear{Landi2005}) considered the nonlinear propagation of a harmonic Alfv\'en wave train in a 2.5D geometry, consisting of a magnetic X-point threaded by an axial magnetic field. Interestingly, the authors find that the driven Alfv\'en waves couple to the fast mode through the magnetic geometry, and that the generated fast waves have a frequency equal to that of the driven Alfv\'en waves and an amplitude that scales linearly with the amplitude of the incoming Alfv\'en waves. This is different to nonlinear formation of fast waves from a propagating Alfv\'en wave (due to the ponderomotive force), in which the generated fast waves have a frequency {\emph{twice}} that of the driven Alfv\'en wave (Nakariakov {\it{et al.}} \citeyear{Nakariakov1997}; Tsiklauri {\it{et al.}} \citeyear{Tsiklauri2001}; \citeyear{Tsiklauri2002}) and an amplitude related to the square of the driven Alfv\'en wave amplitude. Thus, Landi {\it{et al.}} (\citeyear{Landi2005}) propose that this indicates a mechanism of mode conversion that differs from the standard nonlinear fast wave excitation via the ponderomotive force. The authors also find that these generated fast waves rapidly develop into fast-mode shocks, and thus the wave dissipation is concentrated into thin current structures. However, this is not surprising as the authors consider very large amplitudes for their driven Alfv\'en waves, precisely so that they can observe shock formation within their simulation domain. The authors do not report whether the fast waves experience the refraction effect or whether the Alfv\'en waves are confined to the magnetic fieldlines that originate on.


 McClements {\it{et al.}} (\citeyear{McClements2006}) and  Ben Ayed {\it{et al.}} (\citeyear{BenAyed2009}) consider radial symmetry for their driven waves, and Landi {\it{et al.}} (\citeyear{Landi2005})  considers a large-amplitude, periodic wave train driven simultaneously  on both side boundaries. Hence, it would be interesting to see if the results of all these papers persist if one considers a single wave pulse, as this would allow the transient behaviour to be observed, and also if the waves were  driven on a single side of the numerical domain.



\section{Three-dimensional null points}\label{section:threedimensionalnullpoints}

Finally, let us now review the behaviour of MHD waves in the neighbourhood of 3D null points. However, surprisingly few papers have been written that address this issue, or at least, papers that concentrate on the the transient propagation of the modes.  Most papers have focused on the dynamics of current formation in an attempt to locate regions where reconnection is most likely to occur, rather than on the transient propagation of the MHD waves.

The first study of the dynamics of current formation at 3D null points was performed by Rickard \& Titov (\citeyear{RT1996}). These authors solved the linear, $\beta=0$ MHD equations, and  studied a 3D null point that is axisymmetric about the spine (i.e. a proper 3D null, see \S\ref{section:threeDnulls}). Perturbations were decomposed into azimuthal modes labelled by mode number $m$, and the analysis is performed in cylindrical geometry.

Several different perturbations  are considered by Rickard \& Titov (see their figure 2) and are driven with a velocity pulse (in $v_r, v_\theta, v_z$) on the boundaries (either the radial boundaries or upper/lower boundaries). Numerical simulations show that axisymmetric perturbations ($m=0$) lead to current accumulation along the spine, whilst the $m=1$ mode produces currents in the fan plane and at the null point itself. For $m>1$, there is no current accumulation anywhere along the skeleton.  Rickard \& Titov also noted that in these axisymmetric equilibria, the azimuthal components decouple from the remaining components when $m=0$, and  that in 2D only the $m=0$ mode was associated with producing currents at the null.

The primary aim of this study was to investigate current accumulation, and not to investigate the behaviour of MHD waves around 3D nulls point. Of course, the perturbations investigated correspond to a combination of  Alfv\'en wave and fast magnetoacoustic waves, but the authors do not use this terminology. In addition, the authors drive a combination of  $v_r, v_\theta, v_z$ which generates both wave types, and thus in some cases it is difficult to confirm that current accumulation results from a certain wave-type. However, some behaviours are clear:  a  pure $m=0$ wave pulse corresponds to a torsional Alfv\'en wave, which the authors note is channelled along the equilibrium magnetic fieldlines. Secondly, the authors note that the $m=1$ motions can propagate across magnetic fieldlines, and that a focusing effect is seen in the evolution of  $j_r$ and $j_\theta$ (as we would expect for the fast wave) for certain disturbances.

Thus, Rickard \& Titov (\citeyear{RT1996}) give a tantalising suggestion that our understanding of 2D behaviour of the  fast and Alfv\'en wave transfers to 3D. However, it is not possible to see from their paper if driving a pure Alfv\'en wave results in generation of a fast mode disturbance through the magnetic geometry, or vice versa, as expected from $\S\ref{guide_field}$. The simulations of Rickard \& Titov also utilise reflecting boundary conditions in their cylindrical geometry. In addition, it can be argued that their investigation is actually 2.5D and not a fully 3D experiment. Nevertheless, Rickard \& Titov (\citeyear{RT1996}) is still a landmark paper for the investigation of 3D MHD wave behaviour in the neighbourhood of magnetic null points. It would be interesting to repeat the work of Rickard \& Titov (\citeyear{RT1996}) but to try to excite pure modes and to focus on the resultant transient wave propagation and possible mode conversion.

The work of Rickard \& Titov (\citeyear{RT1996}) has been extended to multiple null point topologies in a series of papers by Galsgaard and co-workers. Galsgaard {\it{et al.}} (\citeyear{KRRN1996}; \citeyear{Klaus1997}) looked at shearing a 3D potential null point pair, with continuous (opposite) shear on two opposite boundaries (parallel to the separator), where the fieldlines were not returned to their original position. This generated a wave pulse that travelled towards the interior of the domain from both directions, and resulted in current accumulation along the separator line with maximum value at the null points.  Galsgaard {\it{et al.}} (\citeyear{GRR1997}) looked at perturbations in 3D magnetic configurations containing a double null point pair connected by a separator. The boundary motions used were very similar to those described above (i.e. shear the boundary and fix). Their experiments showed that the nulls can either accumulate current individually or act together to produce current along the separator. Galsgaard \& Nordlund (\citeyear{GN1997}) found that when a magnetic structure containing eight null points is perturbed, current density accumulates along separator lines.

In all these Galsgaard {\it{et al.}} papers,  the boundary conditions tried to mimic the effect of photospheric footpoint motions by moving the boundary and holding it fixed. As in  Rickard \& Titov (\citeyear{RT1996}), the terminology of MHD waves was not invoked, and the perturbations considered were a combination of MHD waves.


The first investigation specifically looking at MHD wave propagation about a proper 3D null point was reported in Galsgaard {\it{et al.}} (\citeyear{Galsgaard2003}), where the authors looked at a particular type of wave disturbance and solved the nonlinear $\beta = 0$ MHD equations. Galsgaard {\it{et al.}}  (\citeyear{Galsgaard2003}) investigated the effect of rotating the field lines around the spine to generate  a twist wave (essentially a torsional  Alfv\'en wave) and followed its propagation towards the null point. Twists were imposed simultaneously on the upper and lower boundaries, with both the same and opposite vorticities considered. The authors  found that the helical Alfv\'en wave spreads out as it propagates towards the null point and is confined to the magnetic fieldlines it originates on. As the wave approaches the fan plane, the wave spreads out along the diverging fieldlines and produces current accumulation in the fan plane. 

In addition, the authors also observe the generation of a fast-mode wave, and find that this wave focuses and wraps around the null point. The authors suggest that their twist wave is a pure helical Alfv\'en wave, and that  nonlinear effects generate the fast-mode wave (using the ponderomotive mechanism detailed by Nakariakov {\it{et al.}} \citeyear{Nakariakov1997}). However, the authors themselves also note that the fast wave appears to be absent where the boundary driving is slowly increased and only appears when a near-discontinuity in the boundary driving velocity is utilised. Thus, the exact nature of the fast-mode generation is unclear.

Galsgaard {\it{et al.}} (\citeyear{Galsgaard2003}) confirm that several of the key properties of fast and Alfv\'en waves transfer to 3D geometry. However, again their investigation is actually 2.5D and not fully 3D. In addition, closed boundaries were used, and thus waves propagate toward the boundaries and are reflected back into the domain.


The authors also analyse the linear $\beta=0$ MHD equations using the WKB method for their azimuthally-symmetric 3D null, and find good agreement. They find that the equations for fast and Alfv\'en perturbations decouple, although this is not surprising as, due to their choice of  symmetrical geometry, their resultant equations are two-dimensional (since a proper 3D null point is essentially 2D in cylindrical polar coordinates).

{Pontin} \& {Galsgaard} (\citeyear{PG2007}) and {Pontin} {\it{et al.}} (\citeyear{PBG2007}) have performed numerical simulations in which the spine and fan of a proper 3D null point are subjected to rotational and shear perturbations. They found that  rotations of the fan plane lead to current density accumulation about the spine,  and rotations about the spine lead to current sheets  in the fan plane. In addition,  shearing perturbations lead to 3D localised current sheets focused at the null point itself. Again, this is in good agreement with what we may expect for MHD wave behaviour, i.e.  current accumulation at certain parts of the topology.


The first study of MHD waves in the neighbourhood of an improper 3D null point was investigated by McLaughlin {\it{et al.}} (\citeyear{MFH2008}). These authors consider the propagation of the fast and Alfv\'en waves  about the magnetic equilibrium ${\bf{B}}_0=\left[ x,\epsilon y, -\left(\epsilon +1\right)z\right]$, for  both proper ($\epsilon=1$) and improper ($\epsilon>0$, $\epsilon \neq 1$) 3D potential null points. The authors utilise the 3D WKB approximation of the linear $\beta=0$ MHD equations.

McLaughlin {\it{et al.}} (\citeyear{MFH2008}) find that the fast magnetoacoustic wave experiences refraction towards the magnetic null point, and confirms that, as in 2D,  the effect of this refraction is dictated by the Alfv\'en speed profile.  The fast wave, and thus the wave energy, accumulates at the null point. The current build-up is shown to be exponential and the value of the exponent depends upon $\epsilon$.  Thus, as in 2D, there is preferential heating at the null point for the fast wave.

For the Alfv\'en wave, the authors find that the wave propagates along the equilibrium fieldlines and that a fluid element is confined to the fieldline it starts on.  For an Alfv\'en wave generated along the fan-plane, the wave accumulates along the spine. For an Alfv\'en wave generated across the spine,  the value of $\epsilon$ determines where the wave accumulation will occur: either the fan-plane ($\epsilon=1$), along the $x-$axis ($0<\epsilon <1$) or along the $y-$axis ($\epsilon>1$).  Analytical results show that current density builds up  exponentially, leading to preferential heating in these areas.

{{McLaughlin {\it{et al.}} (\citeyear{MFH2008}) also provide a quantitative analysis of the preferential heating/energy release process. They derive an analytical expression for current evolution resulting from  fast wave propagation along the spine (their equation 23) which is  of the form $|{\bf{j}}|=\omega e^{\left(\epsilon +1 \right) t} / \left[{z_0\left(\epsilon +1 \right)}\right]$ (where $\omega$ is the wave frequency and $z_0$ is the starting point on the spine). Furthermore, in order to give an order-of-magnitude estimate, the authors show that for characteristic coronal values ($L=10\:$Mm, $B=10\:$G, $\rho_0=10^{-12}\:$kg$\:$m$^{-3}$) a planar fast wave propagating along the spine will build-up a current of $0.3\:$mA after a time of $t=1$ seconds, and that resistive effects become non-negligible after a time $t=\log{(\omega \lambda^2/\eta})/4\approx 7\:$seconds (where after $7$ seconds, the fast wave has built up a current of $0.87\:$mA  and has travelled a distance of $7.13\:$Mm). The Alfv\'en wave is degenerate with the fast wave along the spine in a $\beta=0$ plasma, and so has identical estimates for these characteristic conditions.}}

However,  McLaughlin {\it{et al.}} (\citeyear{MFH2008}) are unable to reach any conclusions about the  coupling of the fast and Alfv\'en wave types due to the geometry of the magnetic field.  The WKB approximation, in the form  utilised by McLaughlin {\it{et al.}}, does not take into account the coupling of the fast and Alfv\'en wave types due to the geometry of the magnetic field. Under their approximation, the waves actually see the magnetic field as locally uniform.


Thus, a fully 3D model of MHD wave behaviour in the neighbourhood of a general 3D null point has yet to be investigated, where the investigation tracks the propagation and evolution of each MHD wave, and determines the efficiency of mode-conversion due to the magnetic geometry against that due to nonlinear effects.


\section{Conclusions}\label{section:conclusions}

The behaviour of all three MHD wave types; Alfv\'en, fast and slow wave, has been investigated in the neighbourhood of  2D, 2.5D and{{ (to a certain extent) }}3D magnetic null points, in a variety of geometries and under a variety of assumptions. The main conclusions may be summarised as follows:
\begin{itemize}
\item{The linear, fast magnetoacoustic wave behaviour is dictated by the equilibrium fast wave speed profile (i.e. ${\sqrt{v_A^2+c_S^2}}$), which in low-$\beta$ plasmas can be thought of as the equilibrium Alfv\'en-speed profile. The fast wave is guided towards the null point by a refraction effect and wraps around it. The fast wave slows as it approaches the null, leading to a decrease in length scales and thus an increase in current density close to the null point. In a $\beta=0$ plasma, the fast wave cannot cross the null point and the build-up of current is exponential, indicating that dissipation will occur on a timescale related to $\log {\eta}$. Thus, linear fast wave dissipation is very efficient, and {\emph{null points will be locations for preferential heating}}. For $\beta \neq 0$, the fast wave can cross the null point, due to the finite sound speed there, and wave energy can now escape the null point. In this case, there exists two competing phenomena and the dominate effect is determined by the value of the plasma-$\beta$.}
\item{The linear Alfv\'en wave  propagates along the equilibrium fieldlines and a fluid element is confined to the fieldline it starts on. Since the propagation follows the fieldlines, the Alfv\'en wave spreads out as it approaches the diverging null point. In 2D, all the Alfv\'en wave energy accumulates along the separatrices and the current build-up is exponential in time. In 3D, for an Alfv\'en wave generated along the fan-plane, the wave accumulates along the spine and for an Alfv\'en wave generated across the spine,  the value of $\epsilon$ determines where the wave accumulation will occur:  fan-plane ($\epsilon=1$), along the $x-$axis ($0<\epsilon <1$) or along the $y-$axis ($\epsilon>1$). Hence, all the {\emph{Alfv\'en wave energy will be dissipated along the separatrices/ separatrix surfaces}} and these will be the locations for preferential heating.}
\item{The behaviour of the slow wave in the neighbourhood of null points has received the least attention in the literature. The linear slow wave is found to be wave-guided and accumulates along the separatrices. A low-$\beta$ fast wave can generate/convert into both a high-$\beta$ fast and high-$\beta$ slow wave as it crosses the $v_A=c_S$ mode-conversion layer. Such a layer is a natural consequence for a null point emersed in a $\beta \neq 0$ plasma. In fact, the value of $\beta$ grows as $r^{-2}$ close to  a null point.}
\item{The addition of a weak guiding field leads to linear coupling between the fast and Alfv\'en waves in a low-$\beta$ plasma, and thus the propagation of either mode can generate the other. Such a configuration is, of course, no longer a null point, but rather an X-line. However, the nature of mode-coupling for 3D null points is, at this time, uncertain. Fast waves have been shown to be generated by the propagation of the Alfv\'en wave, but it is unclear if this is due to the fieldline geometry, nonlinear coupling or both waves being simultaneous generated by a common driver.}
\item{Results in 2D show that in the  nonlinear regime, the fast magnetoacoustic wave can deform the equilibrium X-point configuration, leading to a cycle of horizontal and vertical current sheets and associated changes in connectivity. Thus, the system exhibits {\emph{oscillatory reconnection}}.}
\item{It is clear that the equilibrium  magnetic field  plays a fundamental role in the  propagation and properties of MHD waves. In general, an arbitrary  disturbance/perturbation will generate all three wave modes and current accumulation could occur at all the null points, and/or along the spine, fan and separators. Thus, the results described in this review all highlight the importance of understanding the magnetic topology in determining  the locations of wave heating.}
\end{itemize}




However, several big questions still remain in this area:
\begin{itemize}
\item{The nature of the coupling of the three modes in 3D needs to be addressed, and the importance of coupling  due to the magnetic geometry verses nonlinear coupling should be investigated.}
\item{The theory of nonlinear fast waves driving oscillatory reconnection should be extended to study more general disturbances, and to investigate how robust the initial findings of McLaughlin {\it{et al.}} (\citeyear{MDHB2009}) are.}
\item{The key results for the linear fast and Alfv\'en wave make clear predictions as to where preferential heating can occur. It would be interesting to see the theoretical models developed with forward modelling (see e.g. Kilmchuk \& Cargill \citeyear{KC2001}; {De Moortel} \& {Bradshaw} \citeyear{ineke2008}) to provide tell-tale observational signatures, and for these synthetic results to be compared with observational data.}
\end{itemize}


{{In conclusion, we have seen that the study of MHD wave behaviour in the neighbourhood of magnetic null points is  a fundamental plasma process, and can provide critical insights into other areas of plasma behaviour including: mode-conversion ($\S\ref{section:mode conversion}$), oscillatory reconnection ($\S\ref{section:nonlinear}$), quasi-periodic pulsations ($\S\ref{QPPs_section}$) and  phase-mixing ($\S\ref{PM_section}$).}}

{{We now know that the corona is full of MHD wave perturbations   (Tomczyk et al. \citeyear{Tomczyk}). We also know that null points are an inevitable consequence of the distributed isolated magnetic flux sources at the photospheric surface, and potential and non-potential field extrapolations suggest that there are always likely to be null points in the corona (see $\S\ref{statistics}$). Thus, these two areas of scientific study (MHD wave behaviour and magnetic topology) will inevitably encounter each other at some point, i.e. MHD waves {\emph{will}} propagate in the neighbourhood of  coronal null points. Thus, MHD wave propagation about magnetic null points  is itself - theoretically -  a fundamental coronal process. }}


{{However, there is as yet no clear observational evidence for MHD wave behaviour in the neighbourhood of coronal null points. In the lead author's opinion, the successful detection of MHD oscillations around coronal null points will require input from two areas: high-spatial/temporal resolution imaging data as well as potential/non-potential extrapolations from co-temporal magnetograms. Two of the instruments onboard the recently launched Solar Dynamics Observatory (SDO) may satisfy these requirements: the {\emph{Atmospheric Imaging Assembly}} (which will provide high-quality imaging data) and the {\emph{Helioseismic and Magnetic Imager}} (which will provide vector magnetograms). Thus, the first detection of MHD waves in the neighbourhood of coronal null points may be reported in the near future.}}


\begin{acknowledgements}
JAM  acknowledges financial assistance from the Leverhulme Trust. IDM is grateful for support through a Royal Society University Research Fellowship.
\end{acknowledgements}



\end{document}